\documentclass[aps,prb,twocolumn,amsmath,amssymb,nofootinbib,superscriptaddress,floatfix,eqsecnum,]{revtex4-1}

\usepackage{amsmath}
\usepackage{amssymb}
\usepackage{amsthm}
\usepackage[pdftex]{color} 
\usepackage{graphicx}% Include figure files
\usepackage{dcolumn} % Align table columns on decimal point
\usepackage{bm} % bold math
\usepackage[hypertex]{hyperref}
\usepackage{longtable}
\usepackage{ulem}   % to strike things out
\newcommand{\bs}[1]{{\boldsymbol{#1}}}

\begin{document}

\title{
Fractional topological liquids with time-reversal symmetry 
and their lattice realization
      }

\author{Titus Neupert} 
\affiliation{
Condensed Matter Theory Group, 
Paul Scherrer Institute, CH-5232 Villigen PSI,
Switzerland
            } 

\author{Luiz Santos} 
\affiliation{
Department of Physics, 
Harvard University, 
17 Oxford Street, 
Cambridge, Massachusetts 02138,
USA
            } 

\author{Shinsei Ryu} 
\affiliation{
Department of Physics, University of California,
Berkeley, California 94720, USA
            } 

\author{Claudio Chamon} 
\affiliation{
Physics Department, 
Boston University, 
Boston, Massachusetts 02215, USA
            } 

\author{Christopher Mudry} 
\affiliation{
Condensed Matter Theory Group, 
Paul Scherrer Institute, CH-5232 Villigen PSI,
Switzerland
            } 

\date{\today}

\begin{abstract}
  We present a class of time-reversal-symmetric fractional topological
  liquid states in two dimensions that support fractionalized
  excitations. These are incompressible liquids made of electrons, for
  which the charge Hall conductance vanishes and the spin Hall
  conductance needs not be quantized. We then analyze the stability of
  edge states in these two-dimensional topological fluids against
  localization by disorder.  We find a $\mathbb{Z}^{\ }_{2}$ stability
  criterion for whether or not there exists a Kramers pair of edge
  modes that is robust against disorder.  We also introduce an
  interacting electronic two-dimensional lattice model based on
  partially filled flattened bands of a $\mathbb{Z}^{\ }_{2}$
  topological band insulator, which we study using numerical exact
  diagonalization.  We show evidence for instances of the
  fractional topological liquid phase as well as for a
  time-reversal symmetry broken phase with a quantized (charge) Hall
  conductance in the phase diagram for this model. 

\end{abstract}

\maketitle

%\tableofcontents

%\newpage

\medskip
\section{
Introduction
        }
\label{sec:intro}

The hallmark of the integer quantum effect (IQHE)
in an open geometry is the localized nature of 
all two-dimensional (bulk) states
while an integer number of chiral edge states freely propagates
along the one-dimensional boundaries%
~\cite{Klitzing80,Laughlin81,Halperin82}.
These chiral 
edge states are immune to the physics of Anderson localization as long as
backward scattering between edge states of opposite chiralities
is negligible%
~\cite{Laughlin81,Halperin82}.

Many-body interactions among electrons can be treated
perturbatively in the IQHE provided the characteristic many-body energy
scale is less than the single-particle gap between Landau levels. 
This is not true anymore if the chemical potential lies within a Landau level
as the non-interacting many-body ground state 
is then macroscopically degenerate.
The lifting of this extensive degeneracy by the many-body interactions is
a non-perturbative effect.
At some ``magic'' filling fractions that deliver the fractional
quantum Hall effect (FQHE)%
~\cite{Tsui82,Stormer83,Laughlin83a,Haldane83},
a screened Coulomb interaction
selects a finitely degenerate family of ground states, each of which
describes a featureless liquid separated from excitations by an energy gap
in a closed geometry. Such a ground state is called an incompressible 
fractional Hall liquid. The FQHE is an example of topological order%
~\cite{Wen90a,Wen90b,Wen90c}.
In an open geometry, there are branches of 
excitations that disperse across the spectral gap of the two-dimensional bulk,
but these excitations are localized along the direction normal to
the boundary while they  propagate freely along the boundary%
~\cite{Wen90c,Wen91a,Wen91b}.
Contrary to the IQHE, 
these excitations need not all share the same chirality. However,
they are nevertheless immune to the physics of Anderson localization 
provided scattering induced by the disorder between distinct edges
in an open geometry is negligible. 

The integer quantum Hall effect (IQHE) is the 
archetype of a two-dimensional topological band insulator.
The two-dimensional $\mathbb{Z}^{\ }_{2}$ 
topological band insulator is a close relative
of the IQHE that occurs in semi-conductors with sufficiently large
spin-orbit coupling but no breaking of time-reversal symmetry%
~\cite{Kane05a,Kane05b,Bernevig06a,Bernevig06b,Konig07}.
As with the IQHE, the smoking gun for the
$\mathbb{Z}^{\ }_{2}$ topological band insulator is the existence
of gapless Kramers degenerate pairs of edge states that are delocalized along
the boundaries of an open geometry as long as disorder-induced
scattering between distinct boundaries is negligible.
In contrast to the IQHE, it is the odd parity in the number of Kramers pairs
of edge states that is robust to the physics of Anderson localization.

A simple example of a two-dimensional $\mathbb{Z}^{\ }_{2}$
topological band insulator can be obtained by putting together two
copies of an IQHE system with opposite chiralities for up and down
spins. For instance, one could take two copies of Haldane's model%
~\cite{Haldane88},
each of which realizes an integer Hall effect on the honeycomb lattice,
but with Hall conductance differing by a sign.
In this case the spin current is conserved, a consequence of the
independent conservation of the up and down currents, and the spin
Hall conductance inherits its quantization from the IQHE of each spin
species. This example thus realizes an integer quantum spin Hall
effect (IQSHE). However, although simple, this example is not generic. The
$\mathbb{Z}^{\ }_{2}$ topological band insulator does not necessarily
have conserved spin currents, let alone quantized responses.

Along the same line of reasoning, two copies of a FQHE system put
together, again with opposite chiralities for up and down particles,
would realize a fractional quantum spin Hall effect
(FQSHE), as proposed by Bernevig and Zhang.%
~\cite{Bernevig06a} 
(See also Refs.~\onlinecite{Freedman04} and \onlinecite{Hansson04}.)
Levin and Stern in Ref.~\onlinecite{Levin09} proposed 
to characterize two-dimensional fractional topological 
liquids supporting the FQSHE by the criterion that
their edge states are stable against disorder 
provided that they do not break time-reversal symmetry spontaneously.

In this paper, we shall \textit{not} impose the condition that
projection about some quantization axis 
of the electron spin from the underlying microscopic model
is a good quantum number.
We will only demand that time-reversal symmetry holds.
We shall thus distinguish the generic cases of fractional topological
liquids with time-reversal symmetry from the special
cases of fractional topological liquids with time-reversal symmetry
\textit{and} with residual spin-1/2 U(1) rotation symmetry.
In the former cases, the electronic spin is not a good quantum number. 
In the latter cases, conservation of spin allows for the FQSHE.

The subclass of incompressible time-reversal-symmetric liquids 
that we construct here is closely related to
Abelian Chern-Simons theories. Other possibilities
that are not discussed in this publication, may include non-Abelian
Chern Simons theories%
~\cite{Frohlich90,Moore91},
or theories that include, additionally, 
conventional local order parameters (Higgs fields)%
~\cite{Ryu09}.

The relevant effective action for the
Abelian Chern-Simons theory is of the form%
~\cite{Wen90b,Wen90c,Wen91a,Wen91b}
\begin{subequations}
\label{eq: def generic 2N Abelian CS}
\begin{equation}
S:=
S^{\ }_{0}
+
S^{\ }_{e}
+
S^{\ }_{s},
\label{eq: def generic 2N Abelian CS a}
\end{equation}
where
\begin{equation}
S^{\ }_{0}:=
-
\int\,\mathrm{d}t\, \mathrm{d}^2\bs{x}\; 
\epsilon^{\mu\nu\rho}\,
\frac{1}{4\pi}
K^{\ }_{ij}\;a^{i}_{\mu}\;\partial^{\ }_{\nu}\,a^{j}_{\rho},
\label{eq: def generic 2N Abelian CS b}
\end{equation}
\begin{equation}
S^{\ }_{e}:=
\int\,\mathrm{d}t\, \mathrm{d}^2\bs{x}\; 
\epsilon^{\mu\nu\rho}\,
\frac{e}{2\pi}\;
Q^{\ }_{i}\,A^{\ }_{\mu}\partial^{\ }_{\nu} \,a^{i}_{\rho},
\label{eq: def generic 2N Abelian CS c}
\end{equation}
and
\begin{equation}
S^{\ }_{s}:=
\int\,\mathrm{d}t\, \mathrm{d}^2\bs{x}\; 
\epsilon^{\mu\nu\rho}\,
\frac{s}{2\pi}\;
S^{\ }_{i}\,B^{\ }_{\mu}\partial^{\ }_{\nu} \,a^{i}_{\rho}.
\label{eq: def generic 2N Abelian CS d}
\end{equation}
The indices $i$ and $j$ run from 1 to $2N$
and any pair thereof labels an integer-valued matrix element
$K^{\ }_{ij}$ of the symmetric and invertible $2N\times2N$ matrix $K$.
The indices $\mu,\nu$, and $\rho$ run from 0 to 2.
They either label the component $x^{\ }_{\mu}$
of the co-ordinates $(t,\bf{x})$
in ($2+1$)-dimensional space and time
or the component
$A^{\ }_{\mu}(t,\bs{x})$
of an external electromagnetic gauge potential,
or the component $B^{\ }_{\mu}(t,\bs{x})$
of an external gauge potential that couples to the spin-1/2 degrees of freedom
along some quantization axis,
or the components of $2N$ flavors of dynamical Chern-Simons
fields $a^{i}_{\mu}(t,\bs{x})$.
The integer-valued component $Q^{\ }_{i}$ 
of the $2N$-dimensional vector $Q$ represents the $i$-th electric charge 
in units of the electronic charge $e$
and obeys the compatibility condition
\begin{equation}
(-)^{Q^{\ }_{i}}=
(-)^{K^{\ }_{ii}},
\end{equation}
\end{subequations}
for any $i=1,\ldots,2N$ in order for bulk quasiparticles
or, in an open geometry, quasiparticles
on edges to obey a consistent statistics.
The integer-valued component $S^{\ }_{i}$ 
of the $2N$-dimensional vector $S$ represents the $i$-th spin charge 
in units of the spin charge $s$ along some conserved quantization axis.
The operation of time reversal 
maps
$A^{\ }_{\mu}(t,\bs{x})$
into
$+g^{\mu\nu}\,A^{\ }_{\nu}(-t,\bs{x})$;
$B^{\ }_{\mu}(t,\bs{x})$
into
$-g^{\mu\nu}\,B^{\ }_{\nu}(-t,\bs{x})$;
$a^{i}_{\mu}(t,\bs{x})$
into
$-g^{\mu\nu}a^{i+N}_{\nu}(-t,\bs{x})$
for
$i=1,\ldots,N$
and vice versa.
Here, $g^{\ }_{\mu\nu}=\mathrm{diag}(+,-,-)$ is the Lorentz metric
in $(2+1)$-dimensional space and time.
We will show that time-reversal symmetry imposes that the matrix
$K$ is of the block form
\begin{subequations}
\label{decomposition K}
\begin{eqnarray}
&&
K=\left(
\begin{matrix}
    \kappa&\;\Delta\\
    \Delta^{\!\mathsf{T}}&-\kappa
\end{matrix}
\right)
\;,
\label{eq:intro-K-matrix-a}
\\
\nonumber\\
&&
\kappa^{\!\mathsf{T}}=\kappa,\quad \Delta^{\!\mathsf{T}}=-\Delta
\;,
\label{eq:intro-K-matrix-b}
\end{eqnarray}
where $\kappa$ and $\Delta$ are $N\times N$ matrices, 
while the integer-charge vectors $Q$ and $S$ are of the block forms
\begin{eqnarray}
&&
Q=
\left(
\begin{matrix}
\varrho\\
\varrho
\end{matrix}
\right),
\qquad
S=
\left(
\begin{matrix}
\varrho\\
-\varrho
\end{matrix}
\right).
\label{eq:intro-K-matrix-c}
\end{eqnarray}
\end{subequations}
 
The $K$ matrix together with the charge vector $Q$ 
and spin vector $S$ 
that characterize the topological field theory with the action%
~(\ref{eq: def generic 2N Abelian CS a})
define the charge filling fraction, a rational number,
\begin{subequations}
\label{eq: def nu}
\begin{equation}
\nu^{\ }_{e}:=
Q^{\mathsf{T}}\;K^{-1}\;Q
\label{eq: def nu a}
\end{equation}
and the spin filling fraction, another rational number,
\begin{equation}
\nu^{\ }_{s}:=
\frac{1}{2}\,
Q^{\mathsf{T}}\;K^{-1}\;S,
\label{eq: def nuspin a}
\end{equation}
respectively.
The block forms
of $K$ and $Q$ 
in Eq.~(\ref{decomposition K})
imply that
\begin{equation}
\nu^{\ }_{e}=0.
\label{eq: def nu b}
\end{equation}
The  ``zero charge filling fraction''~(\ref{eq: def nu b}) 
states nothing but the fact that there is no charge Hall conductance when
time-reversal symmetry holds. On the other hand, time-reversal symmetry
of the action%
~(\ref{eq: def generic 2N Abelian CS a})
is compatible with a non-vanishing FQSHE as measured by the non-vanishing
quantized spin-Hall conductance
\begin{equation}
\sigma^{\ }_{\mathrm{sH}}:= 
\frac{e}{2\pi}\,\times\,
\nu^{\ }_{s}.
\label{eq: def nu spin b}
\end{equation}
\end{subequations}
The origin of the FQSHE in the action%
~(\ref{eq: def generic 2N Abelian CS a})
is the U(1)$\times$U(1) gauge symmetry when 
$(2+1)$-dimensional space and time has the same topology as
a manifold without boundary. We shall always assume that
the U(1) symmetry associated with charge conservation holds
in this paper. However, we shall not do the same with the U(1)
symmetry responsible for the conservation of the ``spin''
quantum number.
 
The special cases of the FQSHE treated in
Refs.~\onlinecite{Bernevig06a} and \onlinecite{Levin09} 
correspond to imposing the condition
\begin{equation}
\Delta=0
\end{equation}
on the $K$ matrix in Eq.~(\ref{eq:intro-K-matrix-a}).
This restriction is, however, not necessary 
to treat either the FQSHE or the generic case when there is
no residual spin-1/2 U(1) symmetry in the underlying microscopic
model.

The effective topological field theory%
~(\ref{eq: def generic 2N Abelian CS}) 
with the condition for time-reversal symmetry%
~(\ref{decomposition K})
is made of $2N$ Abelian Chern-Simons fields.
As is the case with the FQHE, when two-dimensional space is a
manifold of genus one without boundary (i.e., when two-dimensional space
is topologically equivalent to a torus),
it is characterized by distinct topological sectors%
~\cite{Wen90a,Wen90b,Wen90c}.
All topological sectors are in one-to-one correspondence 
with a finite number 
$\mathcal{N}^{\ }_{\mathrm{GS}}$
of topologically degenerate ground states
of the underlying microscopic theory%
~\cite{Wen90a,Wen90b,Wen90c}.
This degeneracy is nothing but the magnitude of the determinant
$K$ in Eq.~(\ref{eq: def generic 2N Abelian CS a}), 
which is, because of the block structure%
~(\ref{eq:intro-K-matrix-a}),
in turn given by 
\begin{eqnarray}
\label{eq:intro-degeneracy}
\mathcal{N}^{\ }_{\mathrm{GS}}&=&
\left|
\mathrm{det}
\begin{pmatrix}
\kappa
&
\Delta
\\
\Delta^{\!\mathsf{T}}
&
-\kappa
\end{pmatrix}
\right|
\nonumber\\
&=&
\left|
\det
\left(
\begin{matrix}
\Delta^{\!\mathsf{T}}
&
-\kappa
\\
\kappa
&
\;\Delta
\end{matrix}
\right)
\right|
\nonumber\\
&=&
\left|
{\rm Pf}
\left(
\begin{matrix}
\Delta^{\!\mathsf{T}}
&
-\kappa\\
\kappa
&
\;\Delta
\end{matrix}
\right)
\right|^2
\nonumber\\
&=&
\left(
\mathrm{integer}
\right)^2.
\end{eqnarray}
To reach the last line we made used of the fact that
the $K$ matrix is integer valued. We thus predict
that the class of two-dimensional time-reversal-symmetric
fractional topological liquids, 
whose universal properties are captured by 
Eqs.~(\ref{eq: def generic 2N Abelian CS}) 
and~(\ref{decomposition K}),
are characterized by a topological ground state degeneracy that
is always the square of an integer, even if $\Delta\ne 0$,
when space is topologically equivalent to a torus. (Notice
that the condition that $\Delta$ is anti-symmetric implies that
non-vanishing $\Delta$ can only occur for $N>1$.)

We discuss in detail the stability of the edge states associated with
the bulk Chern-Simons action
~(\ref{eq: def generic 2N Abelian CS}) 
obeying the condition for the time-reversal symmetry%
~(\ref{decomposition K}).
We consider a single one-dimensional
edge and construct an interacting quantum field theory for 
$1\leq N^{\ }_{\mathrm{K}}\leq N$
pairs of Kramers degenerate electrons subject to
strong disorder that preserves time-reversal symmetry.
[The integer $2N^{\ }_{\mathrm{K}}$ is the number of odd charges
entering the charge vector $Q$ (Ref.~\cite{footnote})].
We identify the conditions under which at least one Kramers
degenerate pair of electrons remains gapless in spite of the
interactions and disorder.  Our approach is here inspired by the
stability analysis of the edge states performed for the single-layer
FQHE by Haldane in Ref.~\onlinecite{Haldane95} (see also Refs.%
~\onlinecite{Kane94} and \onlinecite{Moore98}), by Naud {\it et al.} in
Refs.~\onlinecite{Naud00} and \onlinecite{Naud01} for the bilayer
FQHE, and specially that by Levin and Stern in
Ref.~\onlinecite{Levin09} for the FQSHE. 
As for the FQSHE, our analysis departs from the analysis of
Haldane in that we impose time-reversal symmetry. 
In this paper, we also depart from Ref.~\onlinecite{Levin09}
by considering explicitly the effects of the  off-diagonal elements 
$\Delta$ in the $K$ matrix. Such terms 
are generically present for any realistic
underlying microscopic model independently
of whether this underlying microscopic model supports or not the FQSHE. 
When considering the stability of the edge theory, 
we allow the residual spin-1/2 U(1) symmetry 
responsible for the FQSHE to be broken
by interactions among the edge modes 
or by a disorder potential. Hence, we seek a criterion for
the stability of the edge theory that does not rely on
the existence of a quantized spin Hall conductance in the bulk
as was done in Ref.~\onlinecite{Levin09}.

The stability of the edge states against disorder hinges on whether
the integer
\begin{equation}
R:=
r\,
\varrho^{\mathsf{T}}\,
(\kappa-\Delta)^{-1}\,
\varrho,
\label{eq: def R}
\end{equation}
is odd (stable) or even (unstable). 
The vector $\varrho$ together with the matrices
$\kappa$ and $\Delta$ were defined in Eq.%
~(\ref{decomposition K}).
The integer $r$ is the smallest integer
such that all the $N$ components of the vector 
$r\,(\kappa-\Delta)^{-1}\, \varrho$ are integers. 
We can quickly check a few simple examples. 
First, observe that, in the limit $\Delta=0$, 
we recover the criterion derived 
in Ref.~\onlinecite{Levin09}. 
Second, when we impose a residual spin-1/2 U(1) symmetry
by appropriately restricting the interactions between
edge channels,
$\nu^{\ }_{\uparrow}=
-\nu^{\ }_{\downarrow}=
\varrho^{\mathsf{T}}(\kappa-\Delta)^{-1}\,\varrho$ 
can be interpreted as the Hall conductivity 
$\sigma^{\ }_{\mathrm{xy}}$ in units of $e^{2}/h$
for each of the separately conserved spin components
along the spin quantization axis. 
The integer $r$ has the interpretation of the number of fluxes 
needed to pump a unit of charge, or the inverse of the ``minimum charge'' 
of Ref.~\onlinecite{Levin09}. Further restricting to the case when
$\kappa=\openone^{\ }_{N}$ gives $R=N$ (i.e., we
have recovered the same criterion as for the two-dimensional
non-interacting $\mathbb{Z}^{\ }_{2}$ topological band insulator).

When there is no residual spin-$1/2$ U(1) symmetry, 
one can no longer relate the index
$R$ to a physical spin Hall conductance. 
Nevertheless, the index $R$ defined in
Eq.~(\ref{eq: def R}) discriminates in all cases whether there is or
not a remaining branch of gapless modes dispersing along the edge.

Before we turn to the detailed analysis of the stability of 
the edge theory in Sec.%
~\ref{sec: Quantum chiral edge theory with time-reversal symmetry},
we shall first pose and answer the question 
of whether one
can realize examples of the Abelian Chern-Simons subclass of
time-reversal-symmetric topological spin states 
in a two-dimensional lattice model in 
Sec.~\ref{sec: Exact diagonalization study of ...}. 
We construct extensions of the lattice models studied in
Refs.~\onlinecite{Neupert11},
\onlinecite{Sheng11}, 
\onlinecite{Wang11a},
and 
\onlinecite{Regnault11}
for which a FQHE was found by
partially filling flat bands with non-trivial Chern numbers,
as proposed in Refs.~\onlinecite{Neupert11},
\onlinecite{Tang11}, and \onlinecite{Sun11}.
(See also Ref.~\onlinecite{Green10} for a discussion
of isolated flat bands with broken time-reversal symmetry in
two-dimensional lattices;
Refs.~\onlinecite{Qi11} and
\onlinecite{Parameswaran11}
for recent progress on the
understanding of the relations between Chern and Landau bands;
and Ref.~\onlinecite{Xiao11}
for predicting that materials 
belonging to the family of heterostructures of transition-metal oxides,
say LaAuO$^{\ }_{3}$,
might realize time-reversal symmetric
topologically non-trivial band insulators with nearly flat bands.)

The systems studied here start with flat bands that realize at
half-filling a two-dimensional integer quantum spin Hall band
insulator. We study with the help of exact diagonalization the nature
of the ground state selected by interactions at partial filling 2/3 of
the lowest band. We find supporting evidences for a featureless ground
state that is consistent with the existence of a spectral gap and a
topological degeneracy $3^2$ in the thermodynamic limit, associated
with a $\Delta=0$ state (i.e., a FQSHE driven by interactions in a
region of the phase diagram). This state is unstable to
spontaneous symmetry breaking of time-reversal symmetry induced by
sufficiently strong interactions, which lands the system onto a
state with degeneracy $3$ (not the square of an integer), which we
identify with a 1/3 FQHE of a magnetized state. 

We close this paper with a summary in Sec.%
~\ref{sec: Summary}. 
We also include two appendices to render the paper
reasonably self-contained.

\medskip
\section{
Exact diagonalization study of a two-dimensional lattice model
with time-reversal symmetry
        }
\label{sec: Exact diagonalization study of ...}

\begin{figure}
\includegraphics[width=0.48\textwidth]{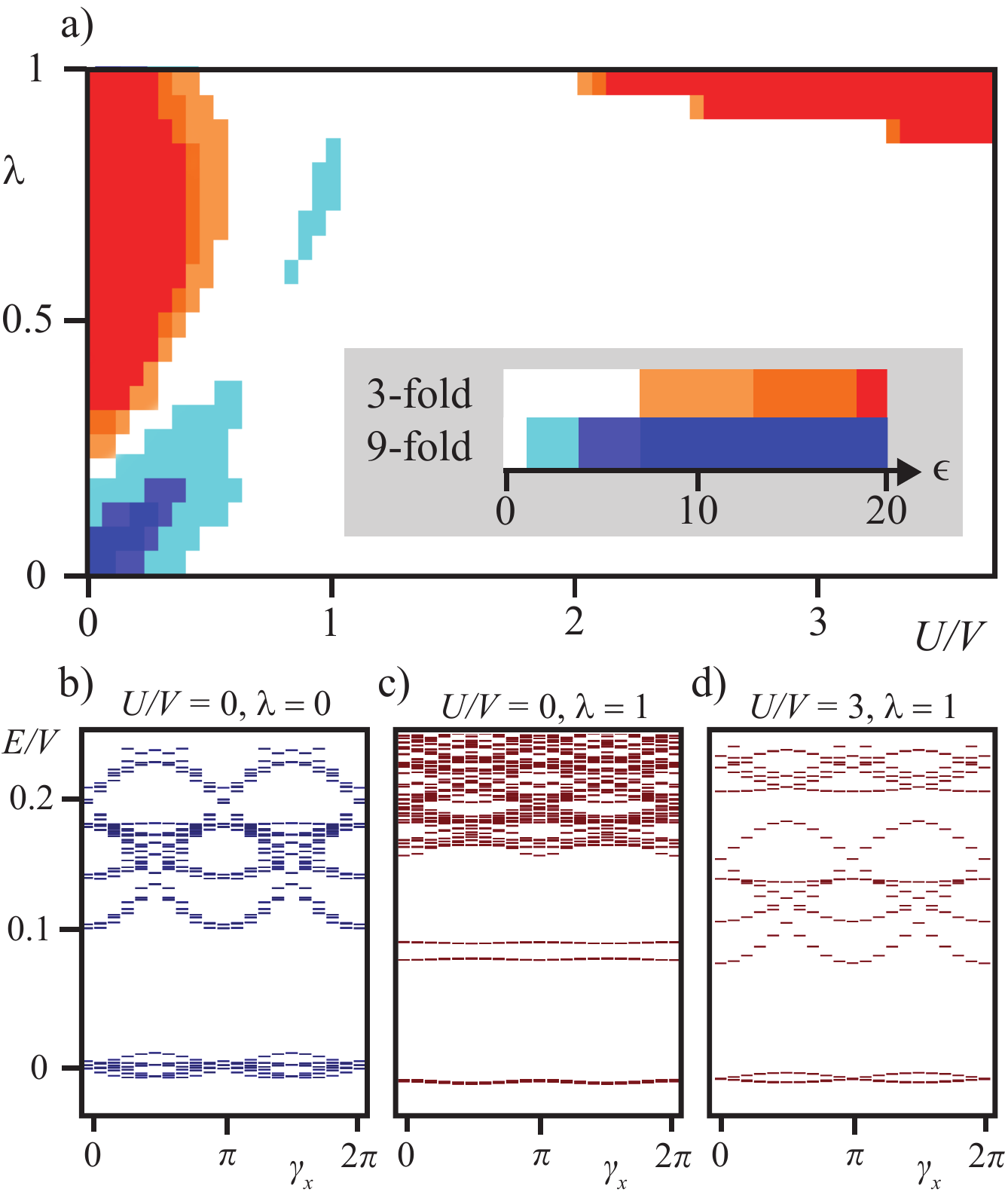}
\caption{
(Color online)
Numerical exact diagonalization results of Hamiltonian%
~\eqref{eq:H lattice} 
for 16 electrons when sublattice A is made of $3\times4$ sites 
and with $t^{\ }_2/t^{\ }_1=0.4$. 
(a) Ground state degeneracies. 
Denote with $E^{\ }_{n}$ the $n$-th lowest energy eigenvalue 
of the many-body spectrum 
where $E^{\ }_{1}$ is the many-body ground state
(i.e., $E^{\ }_{n+1}\geq E^{\ }_{n}$ for $n=1,2,\ldots$).
Define the parameter
$\epsilon^{\ }_{n}:=(E^{\ }_{n+1}-E^{\ }_{n})/(E^{\ }_{n}-E^{\ }_{1})$.
If a large gap 
$E^{\ }_{n+1}-E^{\ }_{n}$
opens up between two consecutive levels
$E^{\ }_{n+1}$ and $E^{\ }_{n}$ 
compared to the cumulative level splitting
$E^{\ }_{n}-E^{\ }_{1}$
between the first $n$ many-body eigenstates induced by
finite-size effects, then the parameter $\epsilon^{\ }_{n}$ is much
larger than unity. The parameter $\epsilon^{\ }_{n}$
has been evaluated for $n=3$ and $n=9$, 
yielding the red and blue regions, respectively. 
For all other $n\neq 1$, no regions with $\epsilon^{\ }_{n}\gtrsim O(1)$ 
of significant size were found. Within the limited range of 
available system sizes, 
it is thus not possible to decide on whether and how the 
level-splitting above the ground state 
in the white regions of the parameter space extrapolates 
in the thermodynamic limit.
(b)--(d) The lowest eigenvalues with spin-dependent twisted boundary 
conditions as a function of the twisting angle $\gamma^{\ }_{x}$. 
The number of low-lying states that are energetically separated 
from the other states is 9, 3, and 3, respectively.
In panel (c), it is the lowest band 
parametrized by $\gamma^{\ }_{x}$ that is three-fold degenerate.
\label{fig: numerics} 
        }
\end{figure}

In Sec.%
~\ref{sec: Quantum chiral edge theory with time-reversal symmetry}, 
we will pose and answer the following question: 
Given a time-reversal-symmetric
incompressible liquid-like ground state 
whose universal properties are encoded
by a low-energy and long-wavelength 
effective quantum field theory for
$2N$ Abelian Chern-Simons fields, 
under what conditions is a Kramers pair of edge modes 
that propagates at the boundary protected against 
Anderson localization as long as time-reversal symmetry is preserved.
In this section, 
we want to address the question, if and
when such a posited topological state emerges in the first place. 
If an incompressible state is connected to a translation invariant band 
insulator, once interactions are switched off adiabatically, 
the answer is entirely governed by the Bloch states of
the single-particle Hamiltonian 
and is well understood~\cite{Schnyder08,Kitaev09,Ryu10}.
If, however, the incompressibility of the state emerges 
from the interactions, the problem is qualitatively different.

As a starting point to study the second situation, 
we shall follow the approach of Refs.~\onlinecite{Bernevig06a} 
and~\onlinecite{Levin09} and consider two decoupled copies 
of the same incompressible fractional quantum Hall (FQH) state 
to compose an incompressible  
time-reversal symmetric fractional quantum spin Hall
(FQSH) state from them. 
Then, we know that the ground state of the system 
is the direct product of the two FQH states and thus 
it is also an incompressible liquid.
It is then natural to ask how stable this state is 
when the two FQH states are coupled, 
that is, whether interactions \emph{between} 
the two FQH states are 
(i) destroying the incompressibility, 
(ii) breaking spontaneously time-reversal symmetry, 
or (iii) generating other incompressible time-reversal symmetric 
states that are not captured by the Abelian Chern-Simons theory in 
Eq.~(\ref{eq: def generic 2N Abelian CS}).

We are going to address this question by numerical exact diagonalization 
of an interacting lattice model, where we will find evidence for all 
scenarios (i), (ii), and (iii). 
It is advantageous to consider a system in which the $z$ component 
of the electronic spin is conserved [i.e., with a residual spin-$1/2$ U(1) symmetry]. 
This makes larger system sizes accessible and allows 
to study the model with spin-dependent twisted boundary conditions 
to determine the ground state degeneracy as explained below.
We consider spinfull electrons hopping on the square lattice 
$\Lambda=\mathrm{A}\cup\mathrm{B}$ 
made up of the two sublattices A and B. 
The Hamiltonian
\begin{equation}
H=H^{\ }_0+H^{\ }_{\mathrm{int}},
\label{eq:H lattice}
\end{equation}
decomposes into a quadratic part 
$H^{\ }_0$ and an interacting contribution 
$H^{\ }_{\mathrm{int}}$.

First, let us define $H^{\ }_0$, which consists of two copies of the
$\pi$-flux phase with flat bands that was studied in 
Ref.~\onlinecite{Neupert11}, 
one copy for each spin-$1/2$ species. 
We denote with $c^\dagger_{\bs{k},\alpha,\sigma}$ 
the creation operator for an electron with lattice momentum 
$\bs{k}$ and spin $\sigma=\uparrow, \downarrow$ 
in the sublattice $\alpha=\mathrm{A},\mathrm{B}$ 
and combine them in the sublattice-spinor 
$\psi^\dagger_{\bs{k},\sigma}:=
\left(
c^\dagger_{\bs{k},\mathrm{A},\sigma},
c^\dagger_{\bs{k},\mathrm{B},\sigma}
\right)
$. 
Then, the second quantized single-particle Hamiltonian reads
\begin{subequations}
\begin{equation}
H^{\ }_0:=
\sum_{\bs{k}\in \mathrm{BZ}}
\left(
\psi^\dagger_{\bs{k},\uparrow}
\frac{
\bs{B}^{\ }_{\bs{k}}\cdot\bs{\tau}
     }
     {
\left|\bs{B}^{\ }_{\bs{k}}\right|
     }
\psi^{\ }_{\bs{k},\uparrow}
+
\psi^\dagger_{\bs{k},\downarrow}
\frac{
\bs{B}^{\ }_{-\bs{k}}\cdot\bs{\tau}^{\mathsf{T}}
     }
     {
\left|\bs{B}^{\ }_{-\bs{k}}\right|
     }
\psi^{\ }_{\bs{k},\downarrow}
\right),
\label{eq: lattice H0}
\end{equation}
where the three-vector $\bs{B}^{\ }_{\bs{k}}$ is defined by
\begin{eqnarray}
&&
B^{\ }_{0,\bs{k}}:=
0,
\\
&&
B^{\ }_{1,\bs{k}}
+
\text{i}
B^{\ }_{2,\bs{k}}
:=
t^{\ }_{1}\,e^{-\text{i}\pi/4}
\left(
1
+
e^{
+\text{i}\left(k^{\ }_{y}-k^{\ }_{x}\right)
  }
\right)
\nonumber\\
&&
\hphantom{
B^{\ }_{1,\bs{k}}
+
\text{i}
B^{\ }_{2,\bs{k}}
:=
         }
+
t^{\ }_{1}\,e^{+\text{i}\pi/4}
\left(
e^{
-\text{i}k^{\ }_{x}
  }
+
e^{
+\text{i}k^{\ }_{y}
  }
\right),
\\
&&
B^{\ }_{3,\bs{k}}
:=
2t^{\ }_{2}
\left(\cos k^{\ }_{x}-\cos k^{\ }_{y}\right),
\end{eqnarray}
\end{subequations}
and the three Pauli matrices 
$\bs{\tau}=(\tau^{\ }_1,\tau^{\ }_2, \tau^{\ }_3)$
act on the sublattice index.
Here, $t^{\ }_1$ and $t^{\ }_{2}$ 
represent the nearest neighbor (NN) and next-nearest neighbor (NNN) 
hopping amplitudes.
The Hamiltonian~\eqref{eq: lattice H0} 
is only well defined if $t^{\ }_{2}\neq0$. 

One verifies that $H^{\ }_{0}$
is both time-reversal symmetric and invariant under 
spin-$1/2$ U(1) rotations. Indeed,
the time-reversal operation $\mathcal{T}$ 
acts on numbers as complex conjugation and on the electron-operators as 
\begin{equation}
\psi^{\dag}_{\bs{k},\uparrow}
\stackrel{\mathcal{T}}{\longrightarrow}
+\psi^{\dag}_{-\bs{k},\downarrow},
\qquad
\psi^{\dag}_{\bs{k},\downarrow}
\stackrel{\mathcal{T}}{\longrightarrow}
-\psi^\dagger_{-\bs{k},\uparrow}.
\end{equation}
The action of the spin-$1/2$ U(1) rotation $\mathcal{R}^{\ }_{\gamma}$
by the angle $0\leq\gamma<2\pi$ is given by
\begin{equation}
\psi^\dagger_{\bs{k},\uparrow}
\stackrel{\mathcal{R}^{\ }_{\gamma}}{\longrightarrow}
e^{+\mathrm{i}\gamma}\psi^\dagger_{\bs{k},\uparrow},
\qquad
\psi^\dagger_{\bs{k},\downarrow}
\stackrel{\mathcal{R}_\gamma}{\longrightarrow}
e^{-\mathrm{i}\gamma}\psi^\dagger_{\bs{k},\downarrow}.
\end{equation}

The spectrum of $H^{\ }_0$ is gaped and comprises  
four dispersionless bands with the energy eigenvalues
\begin{equation}
\varepsilon_{\bs{k},\sigma, \pm}=\pm1,\qquad \sigma=\uparrow, \downarrow.
\end{equation}
Denoting the corresponding single-particle eigenstates by 
$\chi^{\ }_{\bs{k},\sigma,\pm},\ \sigma=\uparrow,\downarrow$, 
we can define the spin-resolved first Chern number
for each of the two pairs of degenerate single-particle bands as
\begin{equation}
\begin{split}
C^{\ }_{\mathrm{s},\pm}:=\frac{1}{2}
\int\limits_{\bs{k}\in \mathrm{BZ}}
\frac{\mathrm{d}^2\bs{k}}{2\pi\mathrm{i}}
\bs{\nabla}^{\ }_{\bs{k}}
\wedge
\bigl(&
\chi^\dagger_{\bs{k},\uparrow,\pm}
\bs{\nabla}^{\ }_{\bs{k}}
\chi^{\ }_{\bs{k},\uparrow,\pm}\\
&
-\chi^\dagger_{\bs{k},\downarrow,\pm}
\bs{\nabla}^{\ }_{\bs{k}}
\chi^{\ }_{\bs{k},\downarrow,\pm}
\bigr).
\end{split}
\end{equation}
We find $C^{\ }_{\mathrm{s},\pm}=\pm 1$.
As a consequence, the non-interacting model exhibits an IQSHE with 
spin-Hall conductivity 
$\sigma^{\mathrm{SH}}_{\mathrm{xy}}=2\times e/(4\pi)$ 
if the chemical potential lies in the single-particle spectral gap.

The repulsive interactions in this model are defined by
\begin{equation}
\begin{split}
H^{\ }_{\mathrm{int}}
:=&\,
U
\sum_{i\in \Lambda}
\rho^{\ }_{i,\uparrow}
\rho^{\ }_{i,\downarrow}
+
V
\sum_{\left\langle ij\right\rangle\in \Lambda}
\Big[
\rho^{\ }_{i,\uparrow}
\rho^{\ }_{j,\uparrow}
+
\rho^{\ }_{i,\downarrow}
\rho^{\ }_{j,\downarrow}
\\
&\,+
\lambda
\left(
\rho^{\ }_{i,\uparrow}
\rho^{\ }_{j,\downarrow}
+
\rho^{\ }_{i,\downarrow}
\rho^{\ }_{j,\uparrow}
\right)
\Big],
\end{split}
\end{equation}
where $\left\langle ij\right\rangle$ 
are directed NN bonds of the square lattice 
$\Lambda=\mathrm{A}\cup\mathrm{B}$, 
$\rho^{\ }_{i,\sigma}$ 
is the occupation number of the site $i$ with electrons of spin 
$\sigma$. The interacting Hamiltonian
$H^{\ }_{\mathrm{int}}$ 
comprises an on-site Hubbard term with the coupling
$U\geq0$ and a NN term which is parametrized by the coupling
$V>0$ and the dimensionless number $\lambda\in[0,1]$. 
The value $\lambda=1$ corresponds to the spin-1/2 SU(2)-symmetric
limit, while all other values of $\lambda$
correspond to the spin-1/2 U(1)-symmetric limit.
These interactions lift the macroscopic degeneracy of 
the single-particle bands. They couple the spin-up and the spin-down sectors, 
if at least one of $U$ or $\lambda$ is non-vanishing.  
Notice that $H^{\ }_{\mathrm{int}}$ shares both the time-reversal and 
spin-$1/2$ U(1) symmetries of the single-particle Hamiltonian 
$H^{\ }_{0}$. 

Periodic boundary conditions are imposed on lattice $\Lambda$ 
whereby sublattice A contains
$L^{\ }_{x}\times L^{\ }_{y}$ sites. 
We fix the number of electrons to be 16 while 
$L^{\ }_{x}=3$ and $L^{\ }_{y}=4$.
We define the filling fraction $2/3$ to be the number of particles, 16,
divided by the number of Bloch single-particle states in the lowest
spin-degenerate band, $(3\times4\times2\times2)/2=48/2=24$.
We then project Hamiltonian~\eqref{eq:H lattice}  
onto the states in the two lower single-particle bands 
$\varepsilon_{\bs{k},\uparrow, -},\ \varepsilon_{\bs{k},\downarrow, -}$, 
thereby
assuming that the single-particle gap is much larger than the energy scale 
of the interactions. Exact diagonalization yields the
many-body spectrum as a function of the
interaction parameters $\lambda$ and $U/V$.
We identify three distinct incompressible states in the 
$\lambda$-$U$ phase diagram [see Fig.~\ref{fig: numerics}(a)].

\textit{Case $\lambda=U/V=0$: decoupled FQH states}. 
The model decouples into two FQH-like states at $2/3$ filling, 
one for each spin orientation. The low-energy effective theory 
for this state could be compatible with the choice
\begin{equation}
K=
\begin{pmatrix}
	+1&+1& 0& 0\\
	+1&-2& 0& 0\\
	 0& 0&-1&-1\\
	 0& 0&-1&+2\\
\end{pmatrix},
\qquad
Q=
\begin{pmatrix}
	1\\
	0\\
	1\\
	0
\end{pmatrix},
\end{equation}
for the $K$ matrix and the charge vector $Q$ in that it
has degeneracy $|\mathrm{det}\,K|=3^2=9$ 
as confirmed by the numerical results.
This phase is destabilized by introducing 
a sufficiently strong coupling between the two FQH states via 
$\lambda$ and $U$. The ability to make a quantitative statement 
on the boundary of this phase in
the phase diagram is here limited by the difficulty in
deciding on the compressibility of a state from extrapolation of 
exact diagonalization studies of small systems sizes 
to the thermodynamic limit.

\textit{Case $\lambda=1$, $U/V>2$: Spontaneous symmetry breaking}. 
We observe that the ground state has the maximal spin-polarization 
that is allowed by the Pauli principle. To interpret this numerical
result, first recall that, after projection onto the lowest bands,
at most
$L^{\ }_{x}\times L^{\ }_{y}$ 
electrons may have the same spin (i.e., 12 for the case at hand). 
Now, the filling fraction is $2/3$ (i.e., 
there are
$4/3\times L^{\ }_{x}\times L^{\ }_{y}=16$ electrons).
If 12 electrons are fully spin polarized, which is what we observe numerically,
then the remaining 
$1/3\times L^{\ }_{x}\times L^{\ }_{y}=4$ 
electrons may form a $1/3$ FQH-like state.
We conjecture that
the low-energy effective theory for this fully spin-polarized
ground state is characterized by the $K$ matrix 
\begin{subequations}
\begin{equation}
K=
\begin{pmatrix}
	+1& 0\\
	 0&-3
\end{pmatrix},
\qquad
Q=
\begin{pmatrix}
	1\\
	1
\end{pmatrix},
\end{equation}
with the filling fraction
\begin{equation}
\nu=
Q^{\mathsf{T}}\,K^{-1}\,Q=2/3.
\end{equation}
\end{subequations}
Clearly, this $K$ matrix does not obey
the decomposition%
~(\ref{decomposition K}), 
since time-reversal symmetry is spontaneously broken. 
The degeneracy $|\mathrm{det}\,K|=3$ is confirmed by the numerical results.
The state thus obtained resembles the conventional double-layer 2/3 FQH state,
with the difference that the electron spins are not fully polarized.

\textit{Case $\lambda=1$, $U/V=0$: Possible paired state}. 
 A time-reversal symmetric state with a spectral gap and a 
three-fold ground state degeneracy is obtained for small $U/V$. 
This state cannot be captured by the Abelian Chern-Simons theory in 
Eq.~(\ref{eq: def generic 2N Abelian CS}), 
since its degeneracy is not the square of an integer, despite 
the time-reversal symmetry. 
One may speculate that this state realizes some real-space pairing of 
spin-up with spin-down electrons, since for small $U/V$ 
it costs little energy to have two electrons of opposite spin 
at the same lattice site.

We close Sec.~\ref{sec: Exact diagonalization study of ...}
with some technical material.
To determine the degeneracy of the ground state unambiguously, 
we have used spin-dependent twisted boundary conditions along 
the $x$ direction defined by 
\begin{equation}
\left\langle 
\bs{r}
+
L^{\ }_{x}
\widehat{\bs{x}}
\left|
\Psi^{\ }_{\gamma^{\ }_{x}}
\right.
\right\rangle=
\left\langle 
\bs{r}
\left|
e^{\mathrm{i}\gamma^{\ }_{x}\sigma^{\ }_3}
\left|
\Psi^{\ }_{\gamma^{\ }_{x}}
\right.
\right.
\right\rangle,
\end{equation}
where $\Psi^{\ }_{\gamma^{\ }_{x}}$ 
is the many-body state, 
$\sigma^{\ }_3$ acts on the spin degrees of freedom, 
and $\widehat{\bs{x}}$ is the corresponding basis vector of the lattice.
Imposing this boundary condition is equivalent to inserting the flux 
$\gamma^{\ }_{x}$ and its time-reversed flux $-\gamma^{\ }_{x}$
for electrons with spin up and spin down quantum numbers, respectively. 
This is a well-defined operation due to the residual spin-$1/2$ U(1) 
symmetry that preserves time-reversal symmetry
for any value of $\gamma^{\ }_{x}$.
For the three cases discussed above, 
Figs.~\ref{fig: numerics}(b)--(d) 
show that the states of the (nearly degenerate)
ground state manifold remain well separated 
from the continuum of states as $\gamma^{\ }_{x}$ 
is varied from 0 to $2\pi$, 
thereby confirming the nine-fold and three-fold degeneracy, respectively.

\begin{figure}
\includegraphics[width=0.4\textwidth]{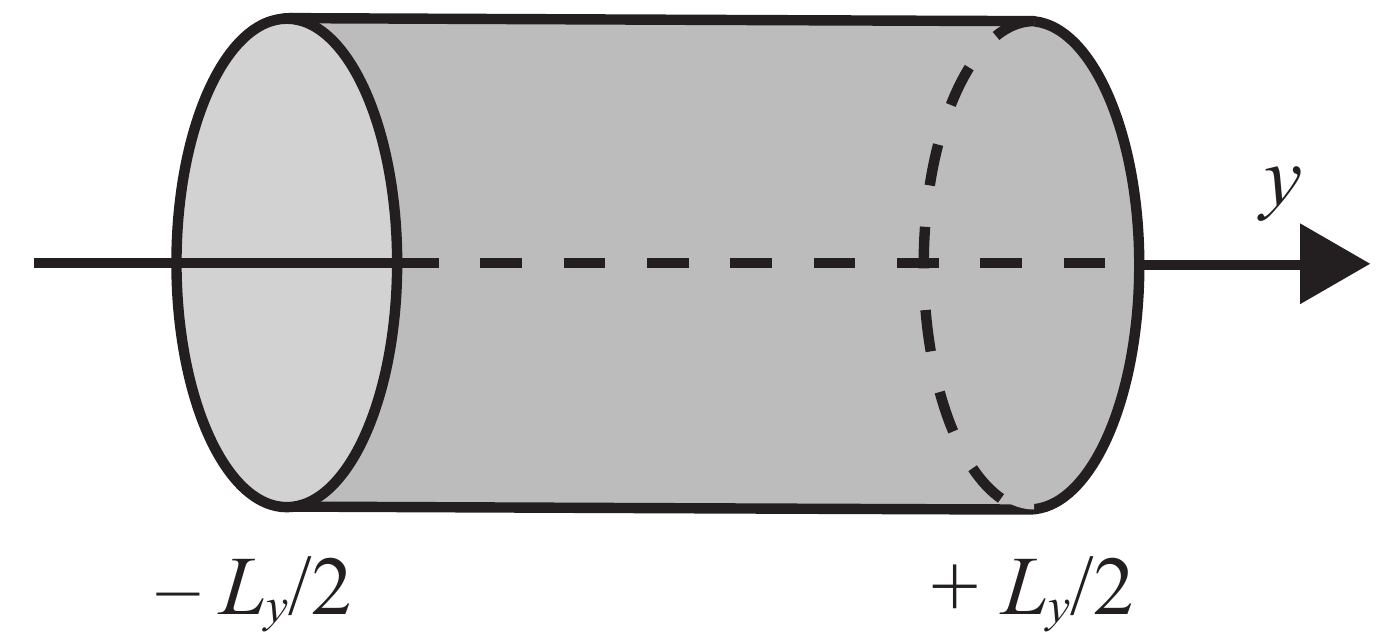}
\caption{
Cylindrical geometry for a two-dimensional band insulator.
The cylinder axis is labeled by the co-ordinate $y$. 
Periodic boundary conditions are
imposed in the transverse direction labeled by the co-ordinate $x$.
There is an edge at $y=-L^{\ }_{y}/2$ and another one at $y=+L^{\ }_{y}/2$.
Bulk states have a support on the shaded surface of the cylinder.
Edge states are confined in the $y$ direction to the vicinity
of the edges $y=\pm L^{\ }_{y}/2$. 
Topological band insulators have the property
that there are edge states freely propagating in the $x$ direction
even in the presence of disorder with the mean free path  
$\ell$ provided the limit $\ell/L^{\ }_{y}\ll 1$ holds.
\label{fig: cylinder geometry} 
        }
\end{figure}

\medskip
\section{
Edge theory with time-reversal symmetry
        }
\label{sec: Quantum chiral edge theory with time-reversal symmetry}

We consider an interacting model for electrons
in a two-dimensional cylindrical geometry
as is depicted in Fig.~\ref{fig: cylinder geometry}.
We demand that (i) charge conservation and time-reversal symmetry 
are the only intrinsic symmetries of the microscopic quantum Hamiltonian,
(ii) neither are broken spontaneously by the many-body ground state,
and (iii), if periodic boundary conditions are assumed
along the $y$ co-ordinate in Fig.~\ref{fig: cylinder geometry},
then there is at most a finite number of
degenerate many-body ground states and each 
many-body ground state is separated from 
its tower of many-body excited states by an energy gap.
Had we relaxed the condition that time-reversal symmetry holds,
the remaining assumptions would be realized for the FQHE.

In the open geometry of Fig.~\ref{fig: cylinder geometry},
the only possible excitations with an energy smaller than the
bulk gap in the closed geometry of a torus
must be localized along the $y$ co-ordinate in the vicinities
of the edges at $\pm L^{\ }_{y}/2$. If $L^{\ }_{y}$ 
is much larger than the characteristic linear extension into the bulk
of edge states,
the two edges decouple from each other.
It is then meaningful to define a low-energy and long-wavelength
quantum field theory for the edge states propagating along 
any one of the two boundaries
in Fig.~\ref{fig: cylinder geometry}, which we take to be of
length $L$ each. 

The low-energy and long-wavelength effective quantum field 
theory for the edge
that we are going to construct is inspired by the construction by
Wen of the chiral Luttinger edge theory for the FQHE%
\cite{Wen91a,Wen90b,Wen91b}.
As for the FQHE, this time-reversal symmetric 
boundary quantum field theory has a correspondence to the effective 
time-reversal symmetric bulk topological quantum-field theory 
built out of $2N$ Abelian Chern-Simon fields and
defined by Eqs.%
~(\ref{eq: def generic 2N Abelian CS}) 
and (\ref{decomposition K})%
~\cite{CS-paper-in-prep}.

The simplest class of quantum Hamiltonians that fulfills requirements
(i)--(iii) can be represented in terms of $2N$ real-valued chiral
scalar quantum fields $\hat{\Phi}^{\ }_{i}(t,x)$ with $i=1,\ldots,2N$
that form the components of the quantum vector field
$\hat{\Phi}(t,x)$. After setting the electric charge $e$, the speed of light
$c$, and $\hbar$ to unity, the Hamiltonian for the system is given by
\begin{subequations}
\label{eq:Mdef quantum edge theory}
\begin{eqnarray}
\hat{H}:=
\hat{H}^{\ }_{0}
+
\hat{H}^{\ }_{\mathrm{int}},
\label{eq:Mdef quantum edge theory a}
\end{eqnarray}
where 
\begin{equation}
\hat{H}^{\ }_{0}:=
\int\limits_{0}^{L}
\mathrm{d}x\,
\frac{1}{4\pi}
\partial^{\ }_{x}\hat{\Phi}^{\mathsf{T}}
\;V\;
\partial^{\ }_{x}\hat{\Phi}
\;,
\label{eq:Mdef quantum edge theory b}
\end{equation}
with $V$ a $2N\times2N$ symmetric and positive definite matrix that
accounts, in this bosonic representation, for the screened
translation-invariant two-body interactions between electrons. The
theory is quantized according to the equal-time commutators
\begin{equation}
\begin{split}
&
\left[
\hat{\Phi}^{\ }_{i}(t,x ),
\hat{\Phi}^{\ }_{j}(t,x')
\right]=
-
\mathrm{i}\pi
\Big(
K^{-1}_{ij}
\;\mathrm{sgn}(x-x')
+
\Theta_{ij}
\Big)
\end{split}
\label{eq: def quantum edge theory e}
\end{equation}
where $K$ is a $2N\times2N$ symmetric and invertible matrix with
integer-valued matrix elements, and the $\Theta$ matrix accounts for
Klein factors that ensure that charged excitations in the theory
(vertex operators) satisfy the proper commutation relations. We review
the construction of the vertex operators in detail 
in Appendix~\ref{app: Chiral bosonic quantum theory}. In
particular, fermionic or bosonic charged excitations are
represented by the normal ordered vertex operators
\begin{equation}
\widehat{\Psi}^{\dag}_{T}(t,x)
:=\ 
:e^{-\mathrm{i}\,T_{i}\,K^{\ }_{ij}\,\hat{\Phi}^{\ }_{j}(t,x)}:
\;,
\label{eq:vertex-def}
\end{equation}
where the integer-valued $2N$-dimensional vector $T$ determines 
the charge (and statistics) of the operator. 
The operator that measures the total charge density is
\begin{equation}
\hat \rho=
\frac{1}{2\pi}\,
Q^{\ }_{i}\;\partial^{\ }_{x}\hat{\Phi}^{\ }_{i}\;,
\label{eq:charge-density-def}
\end{equation}
where the integer-valued $2N$-dimensional charge vector $Q$, 
together with the
$K$-matrix, specify the universal properties of the edge theory. 
The charge $q^{\ }_{T}$ of the vertex operator in
Eq.~(\ref{eq:vertex-def}) follows from its commutation with the charge
density operator in Eq.~(\ref{eq:charge-density-def}), yielding
$q^{\ }_{T}=T^{\mathsf{T}}\,Q$.

Tunneling of electronic charge among the different edge branches is
accounted for by
\begin{equation}
\hat{H}^{\ }_{\mathrm{int}}:=
-
\int\limits_{0}^{L}
\mathrm{d}x
\sum_{T\in\mathbb{L}}
h^{\ }_{T}(x)
:
\cos
\Big(
T^{\mathsf{T}} K\,\hat \Phi(x)
+
\alpha^{\ }_{T}(x)
\Big)
:.
\label{eq:Mdef quantum edge theory c}
\end{equation}
The real functions $h^{\ }_{T}(x)\geq0$ and
$0\leq\alpha^{\ }_{T}(x)\leq2\pi$ encode information about 
the disorder along the edge when position dependent. 
The set 
\begin{equation}
\mathbb{L}:=
\left\{
T\in \mathbb{Z}^{2N}\left|T^{\mathsf{T}} Q=0\right.
\right\},
\label{eq: def mathbb L}
\end{equation} 
encodes all the possible charge neutral
tunneling processes (i.e., those that just rearrange charge among the
branches). This charge neutrality condition implies that the operator 
$\hat{\Psi}^\dagger_T(t,x)$ is bosonic, for it has even charge.
Observe that set $\mathbb{L}$ forms a lattice. 
Consequently, if $T$ belongs to $\mathbb{L}$ so does
$-T$. In turn, relabeling $T$ to $-T$ in 
$\hat{H}^{\ }_{\mathrm{int}}$
implies that
$h^{\ }_{T}(x)=+h^{\ }_{-T}(x)$
whereas
$\alpha^{\ }_{T}(x)=-\alpha^{\ }_{-T}(x)$.
A discussion of the gauge symmetries of this model 
and the properties of $\mathbb{L}$ 
can be found in Appendix~\ref{app: Chiral bosonic quantum theory}.
\end{subequations}

The theory~\eqref{eq:Mdef quantum edge theory}
is inherently encoding interactions:
The terms $\hat{H}^{\ }_0$
and $\hat{H}^{\ }_{\mathrm{int}}$
encode single-particle 
\textit{as well as} many-body interactions
with matrix elements that preserve and break translation symmetry, 
respectively.
Recovering the single-particle kinetic energy
of $N$ Kramers degenerate pairs of electrons
from Eq.~(\ref{eq:Mdef quantum edge theory b})
corresponds to choosing the matrix $V$
to be 
proportional to the unit $2N\times2N$ matrix
with the proportionality constant fixed by the condition
that the scaling dimension of each electron is $1/2$
at the bosonic free-field fixed point defined by
Hamiltonian $\hat{H}^{\ }_{0}$. Of course, to implement
the fermionic statistics for all $2N$ fermions,
one must also demand that all diagonal entries of $K$ are
odd integers in some basis (see Ref.~\onlinecite{footnote}).
While we shall proceed with the interacting bosonic theory here,
we complement this analysis with a review of the stability
of $N$ Kramers pairs of non-interacting fermions presented in 
Appendix%
~\ref{sec: Non-interacting fermionic edge theory with time-reversal symmetry}.

\subsection{
Time-reversal symmetry of the edge theory
           }

The operation of time-reversal on the $\hat \Phi$ fields is defined by
\begin{subequations}
\label{eq:Mdef time reversal trsf}
\begin{equation}
\mathcal{T}\,\hat{\Phi}(t,x)\,\mathcal{T}^{-1}:=
\Sigma^{\ }_{1}\,
\hat{\Phi}(-t,x)
+
\pi
K^{-1}\,
\Sigma^{\ }_{\downarrow}\,
Q,
\label{eq:Mdef time reversal trsf b}
\end{equation}
where
\begin{equation}
\Sigma^{\ }_{1}=
\left(  
\begin{matrix}
0
&
\openone
\\
\openone
&
0
\end{matrix}
\right)
\quad
\mathrm{and}
\quad
\Sigma^{\ }_{\downarrow}=
\left(  
\begin{matrix}
0
&
0
\\
0
&
\openone
\end{matrix}
\right).
\label{eq:sigma-def}
\end{equation}
\end{subequations}
This definition ensures that fermionic and bosonic vertex operators as
in Eq.~(\ref{eq:vertex-def}) are properly transformed under time reversal.
More precisely, one can then construct a pair of fermionic operators
$\hat{\Psi}^{\dag}_{1}$ and $\hat{\Psi}^{\dag}_{2}$ 
of the form~(\ref{eq:vertex-def}) by suitably choosing a pair 
of vectors $T^{\ }_{1}$ and $T^{\ }_{2}$, respectively, in such a way
that the operation of time-reversal maps
$\hat{\Psi}^{\dag}_{1}$ into $+\hat{\Psi}^{\dag}_{2}$
whereas it maps
$\hat{\Psi}^{\dag}_{2}$ into $-\hat{\Psi}^{\dag}_{1}$.
Thus, it is meaningful to interpret
the block structure displayed
in Eq.~(\ref{eq:sigma-def}) 
as arising from the upper or lower projection along some 
spin-1/2 quantization axis.

Time-reversal symmetry on the chiral edge theory%
~(\ref{eq:Mdef quantum edge theory})
demands the following conditions, which we explain below:
\begin{subequations}
\label{eq:Mconditions for TRS on parameters}
\begin{equation}
V=
+\Sigma^{\ }_{1}\,V\,\Sigma^{\ }_{1},
\label{eq:Mconditions for TRS on parameters a}
\end{equation}
\vfill
\begin{equation}
K=
-\Sigma^{\ }_{1}\,K\,\Sigma^{\ }_{1},
\label{eq:Mconditions for TRS on parameters b}
\end{equation}
\vfill
\begin{equation}
Q=
\Sigma^{\ }_{1}\,Q,
\label{eq:Mconditions for TRS on parameters c}
\end{equation}
\vfill
\begin{equation}
h^{\ }_{T}(x)
=h^{\ }_{\Sigma^{\ }_{1}T}(x),
\label{eq:Mconditions for TRS on parameters d}
\end{equation}
\vfill
\begin{equation}
\alpha^{\ }_{T}(x)=
\left(
-\alpha^{\ }_{\Sigma^{\ }_{1}\,T}(x)
+
\pi T^{\mathsf{T}}\,\Sigma^{\ }_{\downarrow}\,Q
\right)
\!\!\!\!
\mod 2\pi.
\label{eq:Mconditions for TRS on parameters e}
\end{equation}
\end{subequations} \medskip
The first two conditions --
Eqs.~(\ref{eq:Mconditions for TRS on parameters a})
and (\ref{eq:Mconditions for TRS on parameters b}) -- 
follow from the requirement that $\hat{H}^{\ }_{0}$ 
be time-reversal invariant. 
In particular, the decomposition~\eqref{decomposition K} of $K$
follows from Eq.~(\ref{eq:Mconditions for TRS on parameters b}) and
$K=K^{\mathsf{T}}$.
The third condition 
--
Eq.~(\ref{eq:Mconditions for TRS on parameters c})
--
states that the charge density is invariant under time reversal. 
In particular, the decomposition~\eqref{decomposition K} of $Q$
follows from Eq.~(\ref{eq:Mconditions for TRS on parameters c}).
Finally, $\mathcal{T} \hat{H}^{\ }_{\mathrm{int}}
\mathcal{T}^{-1}=\hat{H}^{\ }_{\mathrm{int}}$ requires
\begin{widetext}
\begin{equation}
\begin{split}
\sum_{T\in\mathbb{L}}\;
h^{\ }_{T}(x)\,
\cos
\left(
T^{\mathsf{T}}\,K\,\hat{\Phi}(t,x)\vphantom{\Big[}
+
\alpha^{\ }_{T}(x)
\right)
&=
\sum_{T\in\mathbb{L}}\;
\mathcal{T}\, 
\left[
h^{\ }_{T}(x)\,
\cos
\left(
T^{\mathsf{T}}\,K\,\hat{\Phi}(t,x)\vphantom{\Big[}
+
\alpha^{\ }_{T}(x)
\right)
\right]
\mathcal{T}^{-1}
\\
&=
\sum_{T\in\mathbb{L}}\;
h^{\ }_{T}(x)\,
\cos
\left(
-
\left(\Sigma^{\ }_{1}\,T\right)^{\mathsf{T}}
K\,
\hat{\Phi}(-t,x)
+
\alpha^{\ }_{T}(x)
-\pi
T^{\mathsf{T}}\,
\Sigma^{\ }_{\downarrow}\,
Q
\right)
\\
&=
\sum_{T\in\mathbb{L}}\;
h^{\ }_{\Sigma^{\ }_{1}\,T}(x)\,
\cos
\left(
-
T^{\mathsf{T}}
K\,
\hat{\Phi}(-t,x)
+
\alpha^{\ }_{\Sigma^{\ }_{1}\,T}(x)
-\pi
(\Sigma^{\ }_{1}\, T)^{\mathsf{T}}\,
\Sigma^{\ }_{\downarrow}\,
Q
\right)
\\
&=
\sum_{T\in\mathbb{L}}\;
h^{\ }_{\Sigma^{\ }_{1}T}(x)\,
\cos
\left(
T^{\mathsf{T}}
K\,
\hat{\Phi}(-t,x)
-
\alpha^{\ }_{\Sigma^{\ }_{1}\,T}(x)
+\pi
(\Sigma^{\ }_{1} T)^{\mathsf{T}}\,
\Sigma^{\ }_{\downarrow}\,
Q
\right)
\;,
\end{split}
\end{equation}
\end{widetext}
leading to the last two relations
--
Eqs.~(\ref{eq:Mconditions for TRS on parameters d})
and (\ref{eq:Mconditions for TRS on parameters e}) 
--
as the conditions needed to match the two trigonometric expansions.

Disorder parametrized by
$h^{\ }_{T}(x)=+h^{\ }_{-T}(x)$ 
and 
$\alpha^{\ }_{T}(x)=-\alpha^{\ }_{-T}(x)$
and for which the matrix $T$ obeys
\begin{subequations}
\label{eq: T in mathbb{T}--}
\begin{equation}
\Sigma^{\ }_{1}\,T=-T,
\end{equation}
and
\begin{equation}
T^{\mathsf{T}}\,\Sigma^{\ }_{\downarrow}\,Q
\hbox{ is an odd integer},
\end{equation}
\end{subequations}
cannot satisfy the condition%
~(\ref{eq:Mconditions for TRS on parameters e})
for time-reversal symmetry. Such disorder is thus prohibited
to enter $\hat{H}^{\ }_{\mathrm{int}}$ in Eq.%
~(\ref{eq:Mdef quantum edge theory c}), 
for it would break
explicitly time-reversal symmetry otherwise. Moreover, 
we also prohibit any ground state that provides
$\exp\Big(\mathrm{i}T^{\mathsf{T}}\,K\,\hat{\Phi}(t,x)\Big)$
with an expectation value when 
$T$ satisfies Eq.~(\ref{eq: T in mathbb{T}--}), 
for it would break spontaneously 
time-reversal symmetry otherwise.

\medskip
\subsection{Pinning the edge fields with disorder potentials}
\label{subsec:pinning}

Solving the interacting theory%
~(\ref{eq:Mdef quantum edge theory})
is beyond the scope of this paper.
What can be done, however, is to identify those fixed
points of the interacting theory%
~(\ref{eq:Mdef quantum edge theory})
that are pertinent to the question of whether or not some edge modes
remain extended along the edge in the limit of strong disorder
$h^{\ }_{T}(x)\to\infty$ for all tunneling matrices $T\in\mathbb{L}$
entering the interaction%
~(\ref{eq:Mdef quantum edge theory c}) .

This question is related to the one posed and answered by
Haldane in Ref.~\onlinecite{Haldane95} for Abelian FQH states and which,
in the context of this paper, would be as follows:
Given an interaction potential caused by 
\textit{weak} disorder on the edges
as defined by Hamiltonian~\eqref{eq:Mdef quantum edge theory c},
what are the tunneling vectors $T\in\mathbb{L}$ that 
can, in principle, describe relevant perturbations that will cause
the system to flow to a strong coupling fixed point characterized by
$h^{\ }_{T} \rightarrow \infty$
away from the fixed point $\hat{H}^{\ }_{0}$?
(See Ref.~\onlinecite{Xu06}
for an answer to this weak-coupling question
in the context of the IQSHE and
$\mathbb{Z}^{\ }_{2}$ topological band insulators.)
By focusing on the strong coupling limit from the outset, we
avoid the issue of following the renormalization group flow
from weak to strong coupling. Evidently, this point of view
presumes that the strong coupling fixed point is stable
and that no intermediary fixed point prevents it from being reached.

To identify the fixed points of the interacting theory%
~(\ref{eq:Mdef quantum edge theory})
in the strong coupling limit (strong disorder limit)
$h^{\ }_{T}\to\infty$,
we ignore the contribution 
$\hat{H}^{\ }_{0}$ 
and restrict the sum over the tunneling matrices in
$\hat{H}^{\ }_{\mathrm{int}}$ 
to a subset 
$\mathbb{H}$ of $\mathbb{L}$
($\mathbb{H} \subset \mathbb{L}$) 
with a precise definition of $\mathbb{H}$ that will
follow in Eq.~\eqref{eq:def-Hset}.
For any choice of $\mathbb{H}$,
there follows the strong-coupling fixed point Hamiltonian
\begin{equation}
\hat{H}^{\ }_{\mathbb{H}}:=
-
\int\limits_{0}^{L}
\mathrm{d}x
\sum_{T\in\mathbb{H}}
h^{\ }_{T}(x)
:
\cos
\Big(
T^{\mathsf{T}} K\,\hat \Phi(x)
+
\alpha^{\ }_{T}(x)
\Big)
:.
\label{eq: def strong coupling fixed point}
\end{equation}
We conjecture that a fixed point Hamiltonian%
~(\ref{eq: def strong coupling fixed point})
is stable if and only if
the set $\mathbb{H}$ is ``maximal''. 
The study of the renormalization group flows
relating the weak, moderate (if any), and the strong
fixed points in the infinite-dimensional parameter space
spanned by the non-universal data $V$, $h^{\ }_{T}(x)$, and
$\alpha^{\ }_{T}(x)$ is again beyond the scope of this paper.

The reader might wonder why we cannot simply choose
$\mathbb{H}=\mathbb{L}$. This is a consequence of the 
chiral equal-time commutation relations%
~(\ref{eq: def quantum edge theory e}),
as emphasized by Haldane in Ref.~\onlinecite{Haldane95},
that prevent the simultaneous locking of
the phases of all the cosines through
\begin{equation}
\partial_{x}\left(T^{\mathsf{T}}\,K\,\hat{\Phi}(t,x)+\alpha^{\ }_{T}(x)\right)
=
C^{\ }_{T}(x)
\;,
\label{eq: locking condition}
\end{equation}
for some time independent real-valued function $C^{\ }_{T}(x)$.
Even in the strong-coupling limit, there are quantum fluctuations
as a consequence of the  chiral equal-time commutation relations%
~(\ref{eq: def quantum edge theory e})
that prevent minimizing the interaction $\hat{H}^{\ }_{\mathrm{int}}$
by minimizing separately each contribution to the trigonometric
expansion%
~(\ref{eq:Mdef quantum edge theory c}).
Finding the ground state in the strong coupling limit 
is a strongly frustrated problem of optimization.

To construct a maximal set $\mathbb{H}$,
we demand that any $T\in\mathbb{H}$ must satisfy
the locking condition%
~(\ref{eq: locking condition}).
Furthermore, we require that the phases of the cosines 
entering the fixed point Hamiltonian%
~(\ref{eq: def strong coupling fixed point})   
be constants of motion 
\begin{equation}
\label{eq:constant of motion}
\left[
\partial_{x}\left(T^{\mathsf{T}}\,K\,\hat{\Phi}(t,x)\right), \hat{H}_{\mathbb{H}}
\right] = 0\;.	
\end{equation}
To find the tunneling vectors $T\in\mathbb{H}$, we thus need to
consider the following commutator
\begin{widetext}
\begin{equation}
\left[
\partial^{\ }_x\left(T^{\mathsf{T}} K \hat{\Phi}(t,x)\right)
\;,\;
h^{\ }_{T'}(x)\;
\cos\left(T'^{\,\mathsf{T}} K \hat{\Phi}(t,x')+\alpha^{\ }_{T'}(x')\right)
\right]
=
-\mathrm{i}\,2\pi\;
T^{\mathsf{T}} K T^{\prime}\;\,h^{\ }_{T'}(x)\;
\sin\left(T'^{\,\mathsf{T}} K \hat{\Phi}(t,x')+\alpha^{\ }_{T'}(x')\right)
\;
\label{eq: quantum fluctuations between T and T'-eom}
\end{equation}
\end{widetext}
and demand that it vanishes. This is achieved if
$T^{\mathsf{T}}\,K\,T^{\prime}=0$.
Equation~(\ref{eq: quantum fluctuations between T and T'-eom}) implies
that any set $\mathbb{H}$ is composed of the charge neutral
vectors satisfying $T^{\mathsf{T}}\,K\,T'= 0$.
It is by choosing a set $\mathbb{H}$ to be ``maximal'' that
we shall obtain the desired criterion for stability.

\medskip
\subsection{
Stability criterion for edge modes
           }

We presented and briefly discussed in the introduction
(see Sec.~\ref{sec:intro}) 
the criteria for at least one branch of edge
excitations to remain delocalized even in the presence of
strong disorder. Here we prove these criteria. 
The idea is to count the maximum possible number
of edge modes that can be pinned (localized) along the edge
by tunneling processes. 
The set of pinning processes must satisfy
\begin{equation}
T^{\mathsf{T}}\,Q=0
\quad
{\rm and}
\quad
T^{\mathsf{T}}\, K\, T'=0
\;,
\label{eq:def-Hset}
\end{equation}
which defines a set $\mathbb{H}$ introduced in Sec.~\ref{subsec:pinning}.
(Note, however, that $\mathbb H$ is not uniquely determined from 
this condition.)
It is very useful to also
define the real extension $\mathbb V$ of a set $\mathbb H$, by
allowing the tunneling vectors $T$ that satisfy Eq.~(\ref{eq:def-Hset}) to take real
values instead of integer values. Notice that $\mathbb V$ is a vector
space over the real numbers.
We shall also demand that $\mathbb{H}$ forms a lattice that is as dense
as the lattice $\mathbb{L}$ by imposing
\begin{equation}
\mathbb{V}\cap\mathbb{L}=\mathbb{H}.
\end{equation}

For any vector $T\in\mathbb V$, consider the vector $K\,T$. It follows
from Eq.~(\ref{eq:def-Hset}) that $K\,T\perp T', \forall T'\in \mathbb
V$. So $K$ maps the space $\mathbb V$ into an orthogonal space
$\mathbb V^\perp$. Since $K$ is invertible, we have 
$\mathbb{V}^\perp=K\,\mathbb{V}$ as well as 
$\mathbb{V}=K^{-1}\mathbb{V}^\perp$, and
thus $\dim\,\mathbb{V}=\dim\,\mathbb{V}^\perp$. Since 
$\dim\mathbb{V}+\dim\mathbb{V}^\perp\le 2N$, it follows that 
$\dim\mathbb{V}\le N$. 
Therefore (as could be anticipated physically) the maximum number
of Kramers pairs of edge modes that can be pinned is $N$; if that
happens, the edge has no gapless delocalized mode.

Next, let us look at the conditions for which the maximum dimension
$N$ is achieved. If $\dim \mathbb V=\dim \mathbb V^\perp=N$, it
follows that $ \mathbb V\oplus \mathbb V^\perp=\mathbb R^{2N}$,
exhausting the space of available vectors, and thus in this case the
charge vector $Q\in \mathbb V^\perp$ because of
Eq.~(\ref{eq:def-Hset}). Consequently $K^{-1}Q\in\mathbb V$, and we
can construct an integer vector $\bar T\parallel K^{-1}Q$ by scaling
$K^{-1}Q$ by the minimum integer $r$ that accomplishes this (which is
possible because $K^{-1}$ is a matrix with rational entries and $Q$ is
a vector of integers). Such a vector $\bar T\in\mathbb H$ is written as
\begin{equation}
\bar T:=
r
\begin{pmatrix}
+(\kappa-\Delta)^{-1}\,
\varrho\\
-(\kappa-\Delta)^{-1}\,
\varrho
\end{pmatrix}
\;.
\label{eq:Adef T0}
\end{equation}
The existence of $(\kappa-\Delta)^{-1}$ follows from $\mathrm{det}\,K\neq0$
and 
\begin{equation}
\mathrm{det}\, K
=\,
(-)^{N}
\left[\mathrm{det}(\kappa-\Delta)\right]^{2}.
\end{equation}

Using $\bar T$, we construct the integer
\begin{equation}
R:=
-\bar T^{\,\mathsf{T}}\,\Sigma^{\ }_{\downarrow}\,Q
\;,
\label{eq:Adef R}
\end{equation}
which, as we argue below, determines if it is possible or not to 
localize all the modes with the $N$ tunneling operators. Here we employ
Eq.~(\ref{eq:Mconditions for TRS on parameters e}), also noticing
that $\Sigma_1\bar T=-\bar T$, and write
\begin{equation}
\begin{split}
\pi R=&
-\pi \bar T^{\mathsf{T}}\,\Sigma^{\ }_{\downarrow}\,Q\\
&=
\Big(
-\alpha^{\ }_{\bar T}(x)
-\alpha^{\ }_{\Sigma^{\ }_{1}\,\bar T}(x)
\Big)
\!\!\!
\mod 2\pi
\\
&=
\Big(
-\alpha^{\ }_{\bar T}(x)
-\alpha^{\ }_{-\bar T}(x)
\Big)
\!\!\!\mod 2\pi
\\
&
=0
\!\!\!\mod 2\pi
\;,
\label{eq:barT-cond}
\end{split}
\end{equation}
where in the last line we used that 
$\alpha^{\ }_{T}(x)=-\alpha^{\ }_{-T}(x)$ for
all $T\in\mathbb{L}$.
If $\bar{T}$ satisfies Eq.~\eqref{eq:barT-cond},
then $R$ must be an even integer.
If Eq.~\eqref{eq:barT-cond} is violated
(i.e., $R$ is an odd integer)
then $\bar{T}$ is not allowed to enter
$\hat{H}^{\ }_{\mathrm{int}}$ for it would otherwise
break time-reversal symmetry 
[thus $h^{\ }_{\bar{T}}(x)=0$ must always hold in this case 
to prevent $\bar{T}$ from entering $\hat{H}^{\ }_{\mathrm{int}}$].
We therefore arrive at the condition that 

\vspace{0.3cm}
$\bullet$ If the maximum number of edge modes are localized or
gaped, then $R$ must be even.

\vspace{0.3cm}
A corollary is that

\vspace{0.3cm}
$\bullet$ If $R$ is odd, at least one edge branch 
is gapless and delocalized.

\vspace{0.3cm} It remains for us to prove that if $R$ is even, then
one can indeed reach the maximum dimension $N$ for the space of
pinning vectors. This is done by construction. Take all eigenvectors
of $\Sigma_1$ with $+1$ eigenvalue. We can take $(N-1)$ of such vectors,
all those orthogonal to $Q$; for the last one we take $\bar T$. One
can check that these $N$ vectors satisfy Eq.~(\ref{eq:def-Hset}) with
the help of $\Sigma^{\ }_{1}\,K\,\Sigma^{\ }_{1}=-K$ 
[listed in
Eq.~(\ref{eq:Mconditions for TRS on parameters b})] and of 
$\bar T \parallel K^{-1}Q$. Now, the $(N-1)$ vectors 
$\Sigma^{\ }_{1}T=+T$ are of the
form $T^\mathsf{T}=(t^\mathsf{T},t^\mathsf{T})$, where we need to satisfy
$T^{\mathsf{T}}Q=2t^\mathsf{T}\varrho=0$. This leads to 
$T^{\mathsf{T}}\,\Sigma^{\ }_{\downarrow}\,Q$ even, and then
Eq.~(\ref{eq:Mconditions for TRS on parameters e})
brings no further conditions whatsoever. So we can take all these $(N-1)$ 
tunneling vectors. Finally, we take $\bar T$ as constructed above, which is
a legitimate choice since $R$ is assumed even and thus consistent with 
Eq.~(\ref{eq:barT-cond}). Hence, we have constructed the $N$ tunneling vectors
that gap or localize all edge modes, and can state that

\vspace{0.3cm}
$\bullet$ If $R$ is even, then the maximum number of edge modes
are localized or gaped.

As a by-product, we see that it is always possible to
localize along the boundary at least all but one Kramers degenerate pair 
of edge states via the $(N-1)$ tunneling vectors that 
satisfy $\Sigma^{\ }_{1}T=+T$. Thus, either one or no Kramers degenerate
pair of edge state remains delocalized along the boundary 
when translation invariance is strongly broken along
the boundary.

\medskip
\subsection{
The stability criterion for edge modes in the FQSHE
           }
\label{subsec: Stability criterion for the FQSHE}

What is the fate of the stability criterion
when we impose the residual spin-$1/2$ U(1) symmetry
in the model so as to describe an underlying
microscopic model that supports
the FQSHE?
The residual spin-$1/2$ U(1) symmetry is imposed
on the interacting theory%
~(\ref{eq:Mdef quantum edge theory})
by positing the existence of a spin vector 
$S=-\Sigma^{\ }_{1}\,S\in\mathbb{Z}^{2N}$ 
associated to a conserved U(1) spin current. 
This spin vector is the counterpart to the charge vector  
$Q=+\Sigma^{\ }_{1}\,Q\in\mathbb{Z}^{2N}$.
The condition
\begin{subequations}
\label{eq: bosonic version U(1) spin 1/2 symmetry}
\begin{equation}
S=-\Sigma^{\ }_{1}\, S,
\label{eq: compatibility condition on S}
\end{equation}
is required for compatibility with time-reversal symmetry
and is the counterpart to Eq.%
~(\ref{eq:Mconditions for TRS on parameters c}).
Compatibility  with time-reversal symmetry
of $Q$ and $S$ thus implies that they are orthogonal,
$Q^{\mathsf{T}}\,S=0$.
If we restrict the interaction%
~(\ref{eq:Mdef quantum edge theory c})
by demanding that the tunneling matrices obey
\begin{equation}
T^{\mathsf{T}}\,S=0,
\label{eq: T S = 0}
\end{equation}
\end{subequations}
we probe the stability of the FQSHE described by
$\hat{H}^{\ }_{0}$
when perturbed by
$\hat{H}^{\ }_{\mathrm{int}}$
(Ref~\cite{footnote FQSHE}).

To answer this question we supplement the condition 
$T^{\mathsf{T}}Q=0$ on
tunneling vectors that belong to $\mathbb{L}$ and $\mathbb{H}$,
by $T^{\mathsf{T}}S=0$.
By construction, $S$
is orthogonal to $Q$.
Hence, it remains true that 
$\mathbb{H}$
is made of at most $N$ linearly
independent tunneling vectors.

The strategy for establishing the condition for the strong coupling limit of
$\hat{H}^{\ }_{\mathrm{int}}$
to open a mobility gap for all the extended modes of
$\hat{H}^{\ }_{0}$
thus remains to construct the largest
set $\mathbb{H}$ out of as few 
tunneling vectors with $T=-\Sigma^{\ }_1T$
as possible, 
since these tunneling vectors might spontaneously break time-reversal symmetry.

As before, there are $(N-1)$ linearly independent tunneling vectors
with $T=+\Sigma^{\ }_1T$,
while the tunneling matrix
$\bar{T}$ from Eq.~(\ref{eq:Adef T0})
must belong to any $\mathbb{H}$
with $N$ linearly independent tunneling vectors.

At this stage, we need to distinguish the case
\begin{subequations}
\label{eq:Atwo option for T(0) if FQSHE}
\begin{equation}
\bar{T}^{\mathsf{T}}\,S=0,
\label{eq:Atwo option for T(0) if FQSHE a}
\end{equation}
from the case
\begin{equation}
\bar{T}^{\mathsf{T}}\,S\neq 0.
\label{eq:Atwo option for T(0) if FQSHE b}
\end{equation}
\end{subequations}
In the former case, 
the spin neutrality condition%
~(\ref{eq: T S = 0})
holds for $\bar{T}$ and thus
the stability criterion
is unchanged for the FQSHE.
In the latter case,
the spin neutrality condition%
~(\ref{eq: T S = 0}) 
is violated 
so that $\hat{H}^{\ }_{\mathrm{int}}$ is independent
of any tunneling matrix proportional to
$\bar{T}$. Thus, when Eq.%
~(\ref{eq:Atwo option for T(0) if FQSHE b})
holds, as could be the case when
$\kappa\propto\openone^{\ }_{N}$ and $\Delta=0$ say, 
the FQSHE carried by at least one Kramers pair of
edge states of $\hat{H}^{\ }_{0}$
is robust to the strong coupling limit of
the  time-reversal symmetric and residual spin-$1/2$ 
U(1) symmetric
perturbation $\hat{H}^{\ }_{\mathrm{int}}$.

\section{
Summary
        }
\label{sec: Summary}

We have considered in this paper a subclass of 
time-reversal-symmetric fractional topological liquids
without quantized charge and spin Hall conductance.
These states can be viewed
as ``zero filling fraction'' quantum Hall states, 
that are related to an Abelian Chern-Simons bulk theory. 
The states we considered depart from previous constructions 
that place together two copies of
FQHE systems, and as such they do not need to satisfy spin
conservation or display quantized spin Hall conductances.

We have analyzed the stability of the edge theory associated with this
type of state, and obtained a discriminant, the parity of an integer, 
that resolves whether there remains or not delocalized edge states 
in the presence of disorder. 
When the discriminant is even, there are no gapless edge
modes. In contrast, gapless edge modes are protected by time-reversal
symmetry when the discriminant is odd. These results contain as
particular cases those that display a quantized
FQSHE~\cite{Levin09}, where the discriminant has a relation to the
then quantized spin Hall conductance.

We have also presented a concrete lattice realization of a 
FQSHE. There have been numerous studies on the effect of
strong interactions in time-reversal-symmetric systems with
non-trivial topology at half filling% 
~\cite{recent works on interacting band TI}.
In contrast, we have considered the effects of 
interactions at a partial filling of bands.
At 2/3 filling of a lattice with $24$ sites,
exact diagonalization delivers
a ground state with nine-fold degeneracy,
which we interpret as a 
time-reversal-symmetric fractional topological liquid.
We studied the stability of this phase by tuning parameters 
of the electron-electron interaction.
In particular, we found a transition toward a phase of spontaneously 
broken time-reversal symmetry,
which is related to a FQHE at $1/3$ filling. 

Let us remark that the lattice model presented in this paper is an
example of fractionalization in two spatial dimensions without
breaking time-reversal symmetry. It thus joins the ranks with the
other known examples thereof constructed on lattices so far, that of
electron-fractionalization in graphene-like systems in
Refs.~\onlinecite{Hou07,Jackiw07,Chamon08a,Chamon08b} together with the
lattice gauge theory presented in Ref.~\onlinecite{Ryu09}, that of the
triangular lattice quantum dimer model in Ref.~\onlinecite{Moessner01},
and that of the doubled chiral spin liquids in Ref.%
~\onlinecite{Scharfenberger11}.

We would like to close this paper by spotlighting some perspectives on
the differences between the $\mathbb{Z}^{\ }_{2}$ topological
band (weakly interacting) insulators 
and the time-reversal-symmetric fractional topological liquid states 
whose very existences are driven by interactions, 
in particular as to the importance that one should
associate with the bulk and boundary states. In the case of the
non-interacting (weakly interacting)
systems, the edge states play a disproportionally
important role, in that the bulk states are just band insulators
without any ground state degeneracy. On the other hand, the
time-reversal-symmetric fractional topological liquids 
display quite rich bulk phenomena, including the possibility of fractionalized
quasiparticles, regardless of whether gapless edge modes survive or
not. Fractionalized particles can be probed without looking at the
edge: capacitive measurements in the bulk~\cite{Martin04}, for
instance, have revealed fractionalized electrons in the bulk of
$\nu=1/3$ FQH states. From this perspective, 
fractional time-reversal-symmetric topological
liquid states should, as one might expect, be much
richer in content than $\mathbb{Z}^{\ }_{2}$ 
topological band (weakly interacting) insulators.

\medskip
\section*{ACKNOWLEDGMENTS}

We gratefully acknowledge useful discussions with 
Rudolf Morf,
Maurizio Storni,
Xiao-Gang Wen,
and
Andrei Bernevig.
We are indebted to Michael A. Levin
for useful discussions regarding 
the FQSHE.
In particular, we learned from Andrei Bernevig
that he too is studying time-reversal-symmetric
fractional topological liquids.
This work was supported in part by DOE Grant No. DEFG02-06ER46316.
TN and CM thank the
Condensed Matter Theory Visitors Program at Boston
University for support.

\appendix

\section{
Chiral bosonic quantum theory
        }
\label{app: Chiral bosonic quantum theory}

In this Appendix, we review the construction of $2N$ Fermi-Bose 
and $2N$ quasi-particle vertex operators from the chiral bosonic 
quantum fields $\hat{\Phi}^{\ }_{i}(t,x),\ i=1,\ldots,2N,$ that enter
the time-reversal invariant quantum edge theory 
with broken translation invariance%
~(\ref{eq:Mdef quantum edge theory})
and discuss their universal properties.
To this end, we consider the 
\textit{universal data} 
$
\left(
K,Q
\right)
$ 
entering the theory defined by 
Eq.~\eqref{eq:Mdef quantum edge theory} 
as opposed to the \textit{non-universal data} 
$\Big(V,h^{\ }_{T}(x),\alpha^{\ }_{T}(x)\Big)$.

On the chiral bosonic quantum fields we impose the boundary conditions
\begin{equation}
K^{\ }_{ij}\,\hat{\Phi}^{\ }_{j}(t,x+L)=
K^{\ }_{ij}\,\hat{\Phi}^{\ }_{j}(t,x)
+
2\pi n^{\ }_{i}
\label{eq: def quantum edge theory d}
\end{equation}
with $n^{\ }_{i}\in\mathbb{Z}$ for all $i=1,\ldots,2N$. 
Together with the condition that the tunneling vectors $T$ 
are restricted to have integer-valued components,
this ensures that the Hamiltonian $\hat{H}$ is single-valued.

The chiral nature of the bosonic quantum fields
arises from demanding that the equal-time commutator
\begin{equation}
\left[
\hat{\Phi}^{\ }_{i}(t,x ),
\hat{\Phi}^{\ }_{j}(t,x')
\right]=
-
\mathrm{i}\pi
\Big(
K^{-1}_{ij}
\mathrm{sgn}(x-x')
+
\Theta^{\ }_{ij}\Big)
\label{eq: def quantum edge theory e bis}
\end{equation}
holds for any pair $i,j=1,\ldots,2N$.
Here,
\begin{equation}
\Theta^{\ }_{ij}:=K^{-1}_{ik}\,L^{\ }_{kl}\,K^{-1}_{lj}
\label{eq:}
\end{equation}
and the antisymmetric $2N\times2N$ matrix $L$ is defined by
(see Ref.~\onlinecite{Haldane95})
\begin{equation}
L^{\ }_{ij}=
\mathrm{sgn}(i-j)
\left(
K^{\ }_{ij}
+
Q^{\ }_{i}
Q^{\ }_{j}
\right),
\label{eq: def quantum edge theory f}
\end{equation}
where $\mathrm{sgn}(0)=0$ is understood.
It then follows that the quadratic theory%
~(\ref{eq:Mdef quantum edge theory b})
is endowed with chiral equations of motion.
Finally, we need to impose the compatibility conditions
\begin{equation}
(-)^{K^{\ }_{ii}}=
(-)^{Q^{\ }_{i}},
\qquad
i=1,\ldots,2N,
\label{eq: def quantum edge theory g}
\end{equation}
in order to construct local excitations with well-defined statistics.

Define for any $i=1,\ldots,2N$ 
the pair of normal-ordered vertex operators
\begin{subequations}
\begin{equation}
\widehat{\Psi}^{\dag}_{\text{q-p},i}(t,x):=\
:e^{-\mathrm{i}\hat{\Phi}^{\ }_{i}(t,x)}:,
\end{equation}
and
\begin{equation}
\widehat{\Psi}^{\dag}_{\text{f-b},i}(t,x):=\
:e^{-\mathrm{i}K^{\ }_{ij}\,\hat{\Phi}^{\ }_{j}(t,x)}:,
\end{equation}
\end{subequations}
respectively. For any $i=1,\ldots,2N$,
the quasiparticle vertex operator 
$\widehat{\Psi}^{\dag}_{\text{q-p},i}(t,x)$
is multi valued under the transformation%
~(\ref{eq: def U(1) ... U(1) internal symmetry})
provided $|\mathrm{det}\,K|>1$ 
in contrast to 
the Fermi-Bose  vertex operator
$\widehat{\Psi}^{\dag}_{\text{f-b},i}(t,x)$
which is always single valued under the transformation%
~(\ref{eq: def U(1) ... U(1) internal symmetry}).

For any pair $i,j=1,\cdots,N$,
the equal-time commutator%
~(\ref{eq: def quantum edge theory e})
delivers the identities
\begin{subequations}
\label{eq: flavor resolved Q trsf of Psi q-p}
\begin{equation}
\left[
\widehat{\mathcal{N}}^{\ }_{i},
\widehat{\Psi}^{\dag}_{\text{q-p},j}(t,x)
\right]=
\delta^{\ }_{ij}
\widehat{\Psi}^{\dag}_{\text{q-p},j}(t,x),
\label{eq: flavor resolved Q trsf of Psi q-p a}
\end{equation}
\medskip
\begin{equation}
\left[
\widehat{\mathcal{N}}^{\ }_{i},
\widehat{\Psi}^{\dag}_{\text{f-b},j}(t,x)
\right]=
K^{\ }_{ij}\,
\widehat{\Psi}^{\dag}_{\text{f-b},j}(t,x),
\label{eq: flavor resolved Q trsf of Psi q-p b}
\end{equation}
\end{subequations}
and
\begin{subequations}
\label{eq: flavor resolved Q trsf of Psi f-b}
\begin{equation}
\left[
\widehat{\mathcal{C}}^{\ }_{i},
\widehat{\Psi}^{\dag}_{\text{q-p},j}(t,x)
\right]=
K^{-1}_{ij}
\widehat{\Psi}^{\dag}_{\text{q-p},j}(t,x),
\label{eq: flavor resolved Q trsf of Psi f-b a}
\end{equation}
\medskip
\begin{equation}
\left[
\widehat{\mathcal{C}}^{\ }_{i},
\widehat{\Psi}^{\dag}_{\text{f-b},j}(t,x)
\right]=
\delta^{\ }_{ij}\,
\widehat{\Psi}^{\dag}_{\text{f-b},j}(t,x),
\label{eq: flavor resolved Q trsf of Psi f-b b}
\end{equation}
\end{subequations}
respectively. Here,
the quasiparticle vertex operator 
$\widehat{\Psi}^{\dag}_{\text{q-p},i}(t,x)$
is an eigenstate of the conserved topological number 
operator
[recall Eq.~(\ref{eq: def quantum edge theory d})]
\begin{subequations}
\begin{equation}
\begin{split}
\widehat{\mathcal{N}}^{\ }_{i}:=&\,
\frac{1}{2\pi}
K^{\ }_{ij}
\int\limits_{0}^{L}
\mathrm{d}x\,
\left(\partial^{\ }_{x}\hat{\Phi}^{\ }_{j}\right)(t,x)
\\
=&\,
\frac{1}{2\pi}
K^{\ }_{ij}
\left[
\hat{\Phi}^{\ }_{j}(t,L)
-
\hat{\Phi}^{\ }_{j}(t,0)
\right],
\end{split}
\end{equation}
with eigenvalue one, while
the Fermi-Bose vertex operator
$\widehat{\Psi}^{\dag}_{\text{f-b},i}(t,x)$
is an eigenstate of the conserved operator
\begin{equation}
\widehat{\mathcal{C}}^{\ }_{i}:=
\frac{1}{2\pi}
\left[
\hat{\Phi}^{\ }_{i}(t,L)
-
\hat{\Phi}^{\ }_{i}(t,0)
\right],
\end{equation}
\end{subequations}
with eigenvalue one for any $i=1,\ldots,2N$.

\vspace{4cm}

\begin{widetext}
The permutation statistics obeyed by any pair $i,j=1,\ldots,2N$ of
quasiparticle and Fermi-Bose operators follows from the application
of the Baker-Campbell-Hausdorff formula,
\begin{subequations}
\begin{equation}
\begin{split}
\widehat{\Psi}^{\dag}_{\text{q-p},i}(t,x )\,
\widehat{\Psi}^{\dag}_{\text{q-p},j}(t,x')=&\,
\widehat{\Psi}^{\dag}_{\text{q-p},j}(t,x')\,
\widehat{\Psi}^{\dag}_{\text{q-p},i}(t,x )\,
e^{
-\mathrm{i}\pi\, 
\left[
K^{-1}_{ij}
\mathrm{sgn}(x-x')
+
\Theta^{\ }_{ij}
\right]
  },
\end{split}
\label{eq: flavor-resolved q-p statistics}
\end{equation}
and
\begin{equation}
\begin{split}
\widehat{\Psi}^{\dag}_{\text{f-b},i}(t,x )\,
\widehat{\Psi}^{\dag}_{\text{f-b},j}(t,x')=&\,
\widehat{\Psi}^{\dag}_{\text{f-b},j}(t,x')\,
\widehat{\Psi}^{\dag}_{\text{f-b},i}(t,x )\,
e^{
-\mathrm{i}\pi
\left[
K^{\ }_{ij}
\mathrm{sgn}(x-x')
+
L^{\ }_{ij}
\right]
  },
\end{split}
\label{eq: intermediary flavor-resolved f-b statistics}
\end{equation}
when $x\neq x'$, respectively. We conclude that, for any $x\neq x'$,
demanding that the $2N\times2N$ matrix $K$ and the 
$2N$-component charge vector $Q$ 
are integer-valued together with the compatibility condition
(\ref{eq: def quantum edge theory g})
is required to obtain local excitations carrying
the Fermi-Bose permutation statistics
\begin{equation}
\begin{split}
\widehat{\Psi}^{\dag}_{\text{f-b},i}(t,x )\,
\widehat{\Psi}^{\dag}_{\text{f-b},j}(t,x')=&\,
(-)^{Q^{\ }_{i}\,Q^{\ }_{j}}\,
\widehat{\Psi}^{\dag}_{\text{f-b},j}(t,x')\,
\widehat{\Psi}^{\dag}_{\text{f-b},i}(t,x ).
\end{split}
\label{eq: flavor-resolved f-b statistics}
\end{equation}
\end{subequations}
\end{widetext}

\medskip

Let us now deduce the connection between the charge vector $Q$,
the conserved operators $\widehat{\mathcal{C}}^{\ }_{i}$, 
and the total charge density operator $\widehat{Q}$ 
that follows from integrating Eq.~(\ref{eq:charge-density-def})
along the edge.
The charge vector $Q$ 
enters explicitly the theory after coupling the 
$2N$ chiral scalar fields to
an external vector gauge potential with the components 
$A^{\ }_{0}$
and
$A^{\ }_{1}$
through the minimal coupling.
The minimal coupling consists in replacing
the $x$ derivative by the covariant derivative
\begin{subequations}
\begin{equation}
\partial^{\ }_{x}\hat{\Phi}^{\ }_{i}\to
D^{\ }_{x}\hat{\Phi}^{\ }_{i}:=
\left(
\partial^{\ }_{x}
+
Q^{\ }_{i}
A^{\ }_{1}
\right)\hat{\Phi}^{\ }_{i},
\end{equation}
for $i=1,\ldots,2N$
and adding the contribution
\begin{equation}
\hat{H}^{\ }_{\mathrm{current}}:=
\int\limits_{0}^{L}
\mathrm{d}x\,
\frac{1}{2\pi}
A^{\ }_{0}
\left(Q^{\mathsf{T}}\,D^{\ }_{x}\hat{\Phi}\right),
\end{equation}
on the right-hand side of Eq.~(\ref{eq:Mdef quantum edge theory a}).
The theory is then invariant under the pure U(1) 
electro-magnetic gauge transformation
\begin{equation}
\hat{\Phi}\to 
\hat{\Phi}+Q\chi,
\quad
A^{\ }_{1}\to
A^{\ }_{1}-\partial^{\ }_{x}\chi.
\end{equation}
\end{subequations}
We can now define the total charge operator by
\begin{equation}
\widehat{Q}:=
Q^{\ }_{i}\,\mathcal{C}^{\ }_{i}.
\end{equation}
It follows that the charge associated with the quasiparticle operator
$\widehat{\Psi}^{\dag}_{\text{q-p},i}$ and with the Fermi-Bose operator
$\widehat{\Psi}^{\dag}_{\text{f-b},i}$ is given by $K^{-1}_{ij}Q^{\ }_{j}$ and
$Q^{\ }_{i}$, respectively.

By assumption, the integer-valued $2N\times2N$ matrix $K$
is symmetric and invertible. 
Consequently,
its inverse $K^{-1}$ 
is also symmetric, but its matrix elements are rational numbers
whenever $|\mathrm{det}\,K|>1$.
Observe that the model~\eqref{eq:Mdef quantum edge theory}
is invariant under the transformation
\begin{subequations}
\label{eq: def U(1) ... U(1) internal symmetry}
\begin{equation}
\hat{\Phi}(t,x)\to
\hat{\Phi}(t,x)
+
2\pi\,
T^{\ }_{*}
\label{eq: def U(1) ... U(1) internal symmetry a}
\end{equation}
for any $T^{\ }_{*}\in\mathbb{R}^{2N}$ 
that is independent of space and time and such that
\begin{equation}
T^{\mathsf{T}}\,K\,T^{\ }_{*}\in\mathbb{Z},
\qquad
Q^{\mathsf{T}}T^{\ }_{*}=0,
\label{eq: def U(1) ... U(1) internal symmetry b}
\end{equation}
\end{subequations}
for all tunneling vectors $T\in\mathbb{L}$.
The quantum Hamiltonian%
~(\ref{eq:Mdef quantum edge theory})
thus possesses an emergent  global $\big(\mathrm{U}(1)\big)^{2N}$
symmetry compared to the microscopic model.
The set of all rational-valued vectors
$T^{\ }_{*}$ that satisfy conditions%
~(\ref{eq: def U(1) ... U(1) internal symmetry})
is the lattice $\mathbb{L}^{*}$ dual to 
the  lattice $\mathbb{L}$.
When $|\mathrm{det}\,K|>1$,
the  lattice $\mathbb{L}$ is a sublattice
of the dual lattice $\mathbb{L}^{*}$ that is generated by
the quasiparticles carrying a unit topological charge.
The ground state of Hamiltonian%
~(\ref{eq:Mdef quantum edge theory})
with the periodic boundary conditions corresponding to the geometry
of a torus is then degenerate with the degeneracy
$|\mathrm{det}\,K|>1$,
which is nothing but the volume of the unit cell of
the  lattice $\mathbb{L}$ in units of the unit cell
of the dual lattice $\mathbb{L}^{* }$.

Note that there is no unique way to define the dual
pair $\mathbb{L}$ and $\mathbb{L}^{*}$.
For example, 
we can use 
Eq.~(\ref{eq: flavor resolved Q trsf of Psi q-p a}) 
to define the dual  lattice $\mathbb{L}^{*}$
to be based on the hypercubic lattice $\mathbb{Z}^{2N}$ and,
in turn, 
use Eq.%
~(\ref{eq: def U(1) ... U(1) internal symmetry b})
to construct $\mathbb{L}$.
Alternatively,
we can use 
Eq.~(\ref{eq: flavor resolved Q trsf of Psi f-b b}) 
to define the  lattice $\mathbb{L}$
to be based on the hypercubic lattice $\mathbb{Z}^{2N}$ and,
in turn, 
use Eq.%
~(\ref{eq: def U(1) ... U(1) internal symmetry b})
to construct the dual lattice $\mathbb{L}^{*}$.
Either ways, the ratio between the unit cells
of the lattices 
$\mathbb{L}$ 
and
$\mathbb{L}^{* }$
is $|\mathrm{det}\,K|$.
We have chosen the latter option.

\medskip
\section{
Non-interacting fermionic edge theory with time-reversal symmetry
        }
\label{sec: Non-interacting fermionic edge theory with time-reversal symmetry}

In this Appendix, we review a non-interacting fermionic theory to analyze 
the role of disorder on the edge of a two-dimensional band insulator 
when imposing time-reversal symmetry. We consider a cylindrical geometry
as is depicted in Fig.~\ref{fig: cylinder geometry}. If the length 
$L^{\ }_{y}$
of the cylinder is much larger than 
the characteristic linear extension into the bulk of edge states,
the two edges at $y=\pm L^{\ }_{y}/2$ decouple. 
The low-energy and 
long-wave-length random single-particle Hamiltonian that describes any
one of the two edges, 
each supporting $N$ Kramers degenerate pairs of electrons, is then given by
\begin{subequations}
\label{eq: H for time-reversal symmetry for spin-1/2 fermions}
\begin{equation}
\begin{split}
&
\mathcal{H}(x):=
\mathrm{i}\mathrm{v}^{\ }_{\mathrm{F}}\,
\sigma^{\ }_{3}\otimes\openone^{\ }_{N}\,
\partial^{\ }_{x}
+
\mathcal{W}(x)=
\mathcal{T}^{-1}\,\mathcal{H}(x)\,\mathcal{T},
\\
&
\mathcal{W}(x):=
\sum_{\mu=0}^{3}
\sigma^{\ }_{\mu}\otimes W^{\ }_{\mu}(x)=
\mathcal{T}^{-1}\,\mathcal{W}(x)\,\mathcal{T},
\end{split}
\label{eq: H for time-reversal symmetry for spin-1/2 fermions a}
\end{equation}
where the operation of time-reversal for spin-$1/2$ electrons is
represented by
\begin{equation}
\mathcal{T}:=
\mathrm{i}\sigma^{\ }_{2}\mathsf{K}=
-\mathcal{T}^{\mathsf{T}}
\label{eq: H for time-reversal symmetry for spin-1/2 fermions b}
\end{equation}
($\mathsf{K}$ represents the operation of charge conjugation).
Here, we have introduced the unit $2\times2$ matrix 
$\sigma^{\ }_{0}$ and the three Pauli matrices 
$\sigma^{\ }_{1}$, 
$\sigma^{\ }_{2}$, 
and
$\sigma^{\ }_{3}$.
In view of
Eq.~(\ref{eq: H for time-reversal symmetry for spin-1/2 fermions b}),
$\sigma^{\ }_{1}$,
$\sigma^{\ }_{2}$,
and
$\sigma^{\ }_{3}$
are to be interpreted as
the generators of the spin-$1/2$ algebra of the electrons.
The matrix $\openone^{\ }_{N}$ is the unit $N\times N$ matrix.
The matrix elements of the
$N\times N$ Hermitean matrices
$W^{\ }_{\mu}(x)$ with $\mu=0,1,2,3$ 
must obey
\begin{equation}
\begin{split}
&
W^{\ }_{0}(x)=+
W^{* }_{0}(x),
\qquad
W^{\ }_{1}(x)=-
W^{* }_{1}(x),
\\
&
W^{\ }_{2}(x)=-
W^{* }_{2}(x),
\qquad
W^{\ }_{3}(x)=-
W^{* }_{3}(x),
\end{split}
\end{equation}
for time-reversal symmetry to hold. Hence, they can be taken as
random numbers 
obeying the white-noise and Gaussian distribution of mean
\begin{equation}
\left\langle 
\left(W^{\ }_{\mu}\right)^{\ }_{\mathsf{ij}}(x)
\right\rangle=
v^{\ }_{\mathsf{i}}\,
\delta^{\ }_{\mathsf{ij}}\,
\delta^{\ }_{\mu,3},
\label{eq: H for time-reversal symmetry for spin-1/2 fermions c}
\end{equation}
and co-variance
\begin{equation}
\begin{split}
\left\langle 
\left(W^{\ }_{\mu}\right)^{\ }_{\mathsf{ij}}(x )
\left(W^{* }_{\nu}\right)^{\ }_{\mathsf{kl}}(x')
\right\rangle=&\,
\frac{1}{N\ell}
\left(\vphantom{\sum}
\delta^{\ }_{\mathsf{ik}}\,
\delta^{\ }_{\mathsf{jl}}
-
(-)^{\delta^{\ }_{\mu,0}}
\delta^{\ }_{\mathsf{il}}\,
\delta^{\ }_{\mathsf{jk}}
\right)
\\
&\,
\times
\delta^{\ }_{\mu\nu}\,
\delta\left(x-x'\right),
\end{split}
\label{eq: H for time-reversal symmetry for spin-1/2 fermions d}
\end{equation}
\end{subequations}
with 
$\mathsf{i},\mathsf{j}=1,\ldots,N$
the flavor index that
label the Kramers degenerate pairs of electrons.
The length scale $\ell$ is the mean free path
within the Born approximation. 
The channel $\mu=3$ represents forward scattering and the mean
$v^{\ }_{\mathsf{i}}$ for $\mathsf{i}=1,\ldots,N$ results, 
through a gauge transformation, 
in a flavor-dependent shift of the Fermi velocity 
$\mathrm{v}^{\ }_{\mathrm{F}}$.

It was known from the studies of quasi-one-dimensional wires
in the 1980's that the random potential $\mathcal{W}(x)$
localizes all $N$ Kramers degenerate pairs of electrons when $N$ is even%
~\cite{Beenakker97}.
It was only realized with the seminal work of Ando and Suzuura
in Ref.~\onlinecite{Suzuura02b}
on carbon nanotubes that the case of an odd number $N$ 
is (i) of physical relevance and (ii) only localizes
$(N-1)$ Kramers degenerate pairs of electrons, leaving
one pair delocalized along the edge%
~\cite{Takane04}.
It then took two groundbreaking papers
from Kane and Mele
on graphene with spin-orbit coupling 
to make the deep connection that this absence of Anderson localization
is the essence of a two-dimensional
$\mathbb{Z}^{\ }_{2}$ topological insulator%
~\cite{Kane05a,Kane05b}.

Observe that the integer quantum spin Hall effect 
(IQSHE) can be deduced
from the following special case of Eq.%
~(\ref{eq: H for time-reversal symmetry for spin-1/2 fermions}).
If we demand that the symmetry condition
\begin{subequations}
\label{eq: def single-partilce condition for IQSHE}
\begin{equation}
\sigma^{\ }_{3}\,\mathcal{H}(x)\,\sigma^{\ }_{3}=
\mathcal{H}(x)
\label{eq: U(1) residual symmetry}
\end{equation}
also holds in addition to the time-reversal symmetry in Eq.%
~(\ref{eq: H for time-reversal symmetry for spin-1/2 fermions}).
Condition~(\ref{eq: U(1) residual symmetry})
is then nothing but the residual spin-$1/2$ U(1) symmetry
generated by rotations about the spin-$1/2$ quantization axis
$\sigma^{\ }_{3}$. Imposing the residual spin-$1/2$ U(1) symmetry%
~(\ref{eq: U(1) residual symmetry})
on the disorder potential $\mathcal{W}(x)$ amounts to the restrictions
\begin{equation}
W^{\ }_{1}(x)=W^{\ }_{2}(x)=0,
\label{eq: absence of backward scattering}
\end{equation}
\end{subequations}
for all $x$ along the edge. 
Condition~(\ref{eq: absence of backward scattering})
removes all backward scattering channels from the
single-particle disorder potential $\mathcal{W}(x)$.
The single-particle Hamiltonian $\mathcal{H}(x)$ 
thus decomposes into the
direct sum of two Hamiltonians, each of which realizes
an integer quantum Hall edge, but with opposite quantized
Hall conductivity of magnitude $N$ in units of $e^2/h$.
The difference between these quantized Hall conductivities
is proportional to $N$ and yields the 
quantized spin Hall conductivity $2N$ in units of 
$e/(4\pi)$.
 
In Sec.%
~\ref{sec: Quantum chiral edge theory with time-reversal symmetry},
we consider an interacting effective quantum field theory including
local multi-particle interactions that break translation invariance.
The only allowed \textit{underlying microscopic} symmetries of this 
interacting effective quantum field theory are
charge conservation and time-reversal symmetry.
The time-reversal symmetry
in Sec.%
~\ref{sec: Quantum chiral edge theory with time-reversal symmetry},
in particular Eqs.~\eqref{eq:Mconditions for TRS on parameters d} 
and~\eqref{eq:Mconditions for TRS on parameters e},  
are inherited from the following properties of the
single-particle disorder potential in Eq.%
~(\ref{eq: H for time-reversal symmetry for spin-1/2 fermions a}).
If $A$ is any complex-valued matrix,
denote with $\mathrm{abs}\,(A)$ and $\mathrm{arg}\,(A)$
the matrices with the matrix elements 
given by the absolute values and phases
of the entries in $A$, respectively.
One then verifies that
\begin{equation}
\begin{split}
&
\mathrm{abs}\,\left(\Sigma^{\ }_{1}\,\mathcal{W}(x)\,\Sigma^{\ }_{1}\right)=
\mathrm{abs}\,\mathcal{W}(x),
\\
&
\mathrm{arg}\,\left(\Sigma^{\ }_{1}\,\mathcal{W}(x)\,\Sigma^{\ }_{1}\right)=
\pi \sigma^{\ }_{1}\otimes E
-\mathrm{arg}\,\mathcal{W}(x),
\end{split}
\label{eq: trsf mathcal under conjugation by Sigma1}
\end{equation}
where $E$ is the $N\times N$ matrix with one for all matrix elements.


\begin{thebibliography}{99}
\bibitem{footnote}
More precisely,
to guarantee that there are $N^{\ }_{\mathrm{K}}$ 
Kramers degenerate pairs of
electrons in the theory, we demand that there exists a 
space and time independent transformation
\begin{equation}
O=
+\Sigma^{\ }_{1}\,O\,\Sigma^{\ }_{1}
\in\mathrm{SL}(2N,\mathbb{Z})
\nonumber
\end{equation}
such that
\begin{equation}
K\to O^{\mathsf{T}}\,K\,O,
\qquad
V\to O^{\mathsf{T}}\,V\,O,
\nonumber
\end{equation}
and
\begin{equation}
Q\to O\,Q
\nonumber
\end{equation}
with the transformed charge vector 
containing $2N^{\ }_{\mathrm{K}}$ odd integers.
\bibitem{footnote FQSHE}
It is important to observe that
the quadratic Hamiltonian%
~(\ref{eq:Mdef quantum edge theory b})
has a much larger symmetry group than the
interacting Hamiltonian%
~(\ref{eq:Mdef quantum edge theory c}).
For example, $\hat{H}^{\ }_{0}$
commutes with the transformation
\begin{equation}
\begin{split}
\hat{\Phi}(t,x)
\longrightarrow
\hat{\Phi}(t,x)
+
\pi
K^{-1}\,
\Sigma^{\ }_{\downarrow}\,
S.
\end{split}
\nonumber
\label{eq: def SU(2) spin trsf}
\end{equation}
One verifies that the transformation law
of a Kramers doublet of fermions
under this transformation is the one expected from
a rotation about the quantization axis of
the residual spin-1/2 U(1) symmetry
provided the parities of the components of $S$ are the same
as those of $Q$. Hence, $\hat{H}^{\ }_{0}$ has,
by construction, the residual spin-1/2 U(1) symmetry 
even though a generic
microscopic model with time-reversal symmetry 
does not. This residual spin-1/2 U(1) symmetry of
$\hat{H}^{\ }_{0}$ is broken by
$\hat{H}^{\ }_{\mathrm{int}}$, unless
one imposes the additional constraint%
~(\ref{eq: T S = 0})
on the tunneling matrices $T\in\mathbb{L}$ 
allowed to enter the interacting theory defined in Eq.%
~(\ref{eq:Mdef quantum edge theory}).
\end{thebibliography}
\end{document}